\documentclass[journal]{IEEEtran}
\usepackage{subfigure}
\usepackage{graphicx}
\usepackage[numbers,sort&compress]{natbib}
\usepackage{amsmath}

\usepackage{algorithm} 
\usepackage{algorithmic} 
\usepackage{multirow} 
\usepackage{amsthm}
\usepackage{setspace}
\usepackage{color}

\usepackage{setspace}

\usepackage{graphicx,epstopdf,epsfig,subfigure,stfloats,array}
\graphicspath{{./eps/}{./pictures/}}
\DeclareGraphicsExtensions{.eps,.png,.jpg}
\usepackage{float}

\ifCLASSINFOpdf
\else
\fi
\begin{document}
%
\title{TFDASH: A Fairness, Stability, and Efficiency Aware Rate Control Approach for Multiple Clients over DASH}
%
%
%

\author{Chao Zhou,
        Chia-Wen Lin, \emph{Senior Member}, \emph{IEEE},
        Xinggong Zhang,
        Zongming Guo$^*$
\thanks{Manuscript received Nov. 21, 2016; revised July 2, 2017, and Feb. 12, 2017; accepted October 29, 2017. This work was supported in part by KTP project, a Private Transport Protocol developed by Kuaishou for video Live, VOD, and Real-Time communication. This paper was recommended by Associate Editor Dr. Zhu Li}
\thanks{Chao Zhou is with the Beijing Kuaishou Technology Co., Ltd., Beijing, China. (e-mail: zhouchaoyf@gmail.com)}
\thanks{Chia-Wen Lin is with the Department of Electrical Engineering and the Institute of Communications Engineering, National Tsing Hua University, Hsinchu, Taiwan (e-mail: cwlin@ee.nthu.edu.tw).}
\thanks{Xinggong Zhang and Zongming Guo (corresponding author) are with the Institute of Computer Science \& Technology, Peking University, Beijing, China.}
\thanks{Color versions of one or more of the figures in this paper are available online at http://ieeexplore.ieee.org.}}


%
%

\markboth{IEEE Transactions on Circuits and Systems for Video Technology}%
{Shell \MakeLowercase{\textit{et al.}}: Bare Demo of IEEEtran.cls for Journals}

%



\maketitle

\begin{abstract}
Dynamic adaptive streaming over HTTP (DASH) has recently been widely deployed in the Internet and adopted in the industry. It, however, does not impose any adaptation logic for selecting the quality of video segments requested by clients and suffers from lackluster performance with respect to a number of desirable properties: efficiency, stability, and fairness when multiple players compete for a bottleneck link. In this paper, we propose a throughput-friendly DASH (TFDASH) rate control scheme for video streaming with multiple clients over DASH to well balance the trade-offs among efficiency, stability, and fairness. The core idea behind guaranteeing fairness and high efficiency (bandwidth utilization) is to avoid OFF periods during the downloading process for all clients, i.e., the bandwidth is in perfect-subscription or over-subscription with bandwidth utilization approach to 100\%. We also propose a dual-threshold buffer model to solve the instability problem caused by the above idea. As a result, by integrating these novel components, we also propose a probability-driven rate adaption logic taking into account several key factors that most influence visual quality, including buffer occupancy, video playback quality, video bit-rate switching frequency and amplitude, to guarantee high-quality video streaming. Our experiments evidently demonstrate the superior performance of the proposed method.
\end{abstract}


%
\IEEEpeerreviewmaketitle

\section{Introduction}
\label{Sec:introduction}
\IEEEPARstart{D}{ynamic} adaptive streaming over HTTP (DASH) has been recently widely adopted for providing uninterrupted video streaming services to users with dynamic network conditions and heterogeneous devices \cite{r3}. In contrast to the past RTP/UDP, the use of HTTP over TCP is easy to configure and, in particular, can greatly simplify the traversal of firewalls and network address translators. Besides, the deployment cost of DASH is relatively low since it employs standard HTTP servers and, therefore, can easily be deployed within content delivery networks. In DASH, a video clip is encoded into multiple versions at different bit-rates, each being further divided into small video segments containing seconds or tens of seconds worth of video. At the client side, a DASH client continuously requests and receives video segments from DASH servers that own the segments. To adapt the video bit-rate to a varying network bandwidth, DASH allows clients to request for video segments from different versions of a video, each of which being coded with a specific bit-rate. This is known as dynamic rate adaptation, which is one of the most important features for DASH since it can automatically throttle the visual quality to match the available bandwidth so that a user receives the video at the maximum quality possible.

it is very challenging to provide satisfactory user experience during a whole video session under highly dynamic network conditions. Without an effective rate adaptation algorithm, a DASH client may suffer from frequent interruptions and degraded visual quality. For example, on one hand, a video bit-rate higher than the available bandwidth would cause network congestion. On the other hand, when the video bit-rate is lower than the available bandwidth, the visual quality cannot reach the maximum allowed by the available bandwidth. Besides, during the playback, smooth video bit-rate is preferred, and frequent bit-rate switching is annoying to users \cite{r4}. However, this is contradictory with the characteristics of the available bandwidth which, generally, is time-varying.

Furthermore, when multiple clients are competing over a bottle-neck link, although the DASH clients employ HTTP over TCP which has proven to be fair, their estimated bandwidth can be significantly different which usually leads to several performance problems, such as inefficiency, instability, and unfairness. This phenomenon is mainly caused by the ON-OFF (ON means downloading a segment, and OFF means staying idle) in DASH \cite{akhshabi2012happens}. In Section \ref{subSec:problem statement}, we shall explain these problems in more details.

Although there are quite a few works addressing rate adaptation problems for video transport over DASH, several research problems about video streaming for multiple clients over DASH are still open and challenging. For example, there is a trade-off between the stability and efficiency for video bit-rate adaptation. Generally, under  time-varying bandwidth, requesting a low video bit-rate gives more room for rate selection, and therefore well ensure high stability (smoothness) and continuous video playback. This, however, leads to low video quality with low bandwidth utilization (under-utilization or inefficiency). In addition, there is a trade-off between the sensitivity and the stability. Since channel bandwidth is inherently time-varying, high sensitivity of a rate control approach usually makes the video bit-rate match the bandwidth well, but may lead to high instability. These challenges become more troublesome when multiple clients are competing over a bottleneck link since each client will try to maximize its received video quality without considering the others. Besides, the fairness problem arises for multiple clients over DASH since the DASH client is deployed over HTTP/TCP. In this case, the trade-off among different performance factors becomes more challenging as will be detailed in Section \ref{subSec:problem statement}.

In this work, we propose a Throughput-Friendly DASH (TFDASH) rate control scheme for multiple clients over DASH, to well balance the trade-offs among efficiency, stability, and fairness. First, considering that the estimated bandwidth is generally not equal to the fair shared bandwidth as explained in Section \ref{subSec:problem statement}, the probed bandwidth, which is obtained by our proposed Logarithmic Increase Multiplicative Decrease (LIMD) based bandwidth probing scheme, is adopted to guide the rate adaptation. For the fairness consideration, we focus on preventing the occurrence of OFF phenomenon during the downloading process, this is because only when the bandwidth is under over-subscription, the bandwidths estimated by the clients individually are approximately equal and fairness can therefore be reached. Besides, without OFF periods, i.e., there is no idle period during the downloading process, the bandwidth utilization is naturally high. However, in this case, how to maintain the system stability becomes challenging since the bandwidth is time-varying, as a too high video bit-rate leads to congestion and playback freeze, while a too low video bit-rate causes buffer overflow without OFF periods. Thus, considering that the available video bit-rates are limited and discrete, the actual video bit-rate must be switched up and down around the fairly shared bandwidth, leading to high instability. Therefore, we adopt a dual-threshold buffer model proposed in our previous work \cite{tmm-26, r6} that employ two thresholds (a low threshold for preventing buffer underflow and a high threshold for preventing buffer overflow) to smooth out the video bit-rate. With this model, a DASH client will continuously download video segments without the need of OFF periods. Specifically, if the available bandwidth is higher than the requested video bit-rate, the buffer occupancy will increase and then approach to the high threshold, and when the buffer occupancy exceeds the high threshold,  the requested video bit-rate will be switched up to prevent buffer overflow, and vice versa. Therefore, without OFF periods, the fairness and efficiency problems are resolved. Furthermore, when the buffer occupancy is in between the two thresholds, a probability driven rate adaptation scheme to ensure high visual quality by jointly considering several key factors including buffer occupancy, video playback quality, video bit-rate switching frequency and amplitude.

\section{Background and Motivation}
\label{Sec:related work}
Although DASH is relatively new, due to its popularity, it has attracted much research effort recently. For example, Watson \cite{r5} systematically introduced the DASH framework of Netflix, which has been the largest DASH stream provider in the world.

As mentioned above, dynamic rate adaptation is a key feature of DASH since it can automatically throttle the visual quality to match the available bandwidth so that each user receives the video with the maximum quality possible. Akhshabi \emph{et al}. \cite{tmm-7} compared the rate adaption schemes used for three popular DASH clients: Netflix client \cite{r5}, Microsoft Smooth Streaming \cite{tmm-9}, and Adobe OSMF \cite{tmm-11}. It was reported in \cite{tmm-7} that none of the DASH client-based rate adaptation is good enough, as they are either too aggressive or too conservative. Some clients even just switch between the highest and lowest bitrates. Also, all of them lead to relatively long response time under the shift of network congestion level. Existing rate adaptation schemes for DASH, such as bandwidth-based schemes \cite{tmm-11,tmm-16,tmm-17} and buffer-based schemes \cite{tmm-18,r24}, aim at either achieving a high bandwidth utilization efficiency by dynamically adapting the video bitrate to match an available bandwidth, or maintaining continuous video playback by smoothing the video bitrate to avoid buffer overflow/underflow. Nevertheless, due to unavoidable bandwidth variations, existing schemes usually cannot achieve a good trade-off between video bitrate smoothness and bandwidth utilization. In our previous work \cite{r6}, a control theoretic approach was proposed to achieve a good trade-off between video rate smoothness and bandwidth utilization. In \cite{r6}, a dual-threshold based buffer occupancy model was proposed to smooth out the short-term bandwidth variations so as to maintain the smoothness of video rate. The main objective of the rate adaption method in \cite{r6} is, however, to avoid buffer overflow and underflow without considering other factors, such as switching frequency and amplitude, which also can affect the perceived visual quality \cite{r14}. Furthermore, these rate control algorithms are designed with the underlying assumption that the TCP downloading throughput observed by a client is its fair share of the network bandwidth, and the fairness problem, which seriously affect the performance for multiple clients over DASH, has not been taken into account. However, due to the OFF intervals during the downloading process \cite{akhshabi2012happens}, such schemes often lead to video bit-rate oscillations, under-utilization,  and unfair bandwidth sharing when multiple clients compete over a common bottleneck link, as will be demonstrated in Section \ref{subSec:problem statement}.

Recently, the problems with video streaming for multiple DASH clients competing over a common bottleneck have been studied in \cite{jiang2012improving,zhu2013fixing,jiang2014improving,li2014probe}. In \cite{jiang2012improving,jiang2014improving}, the authors studied bit-rate adaptation problems and identified the causes of several undesirable interactions that arise as a consequence of overlaying the video bit-rate adaptation over HTTP. Further, they  developed a suite of techniques that tried to systematically guide the trade-offs between stability, fairness, and efficiency. However, the approach did not well consider the factors that most influence visual quality, such as video bit-rate switching frequency and amplitude. Besides, it also did not take care of reducing the OFF periods which is the root causes of unfairness and underutilization in bandwidth allocation \cite{akhshabi2012happens,zhu2013fixing,li2014probe}, instead, it adopted a randomized scheduler to determine when to download a new segment, by which the client needs to wait for a short period. In \cite{zhu2013fixing,li2014probe}, a "probe and adapt" principle was proposed for video bit-rate adaptation where "probe" refers to trial increment of the data rate, instead of sending auxiliary piggybacking traffic, and "adapt" means to select a suitable video bit-rate based on the probed bandwidth. Its additive-increase multiplicative-decrease (AIMD) probing mechanism shares similarities with TCP congestion control \cite{jacobson1988congestion}, which leads to a long convergence time and failure in properly tracking the time-varying bandwidth. While for the "adapt" component, the video bit-rate switching decision is mainly dependent on the probed bandwidth without considering the other factors which most influence visual quality, leading to low bandwidth utilization and high instability.

Besides rate adaptation mechanisms for DASH, there are some other works addressing the fairness problem from several different angles \cite{ma2014access,zhangpresto,joseph2014nova,dubinvideo}. The approach in \cite{ma2014access}  aims to achieve proportional fairness at the packet level by implementing weighted fair queuing. However, these approaches are conducted in a concentrated manner, which has limitation in supporting the large-scale multimedia delivery. A new protocol is proposed in the context of multi-server HTTP adaptive streaming, aiming to improve the user experience by providing better fairness, efficiency and stability in \cite{zhangpresto}. In \cite{joseph2014nova}, the authors jointly optimized the network resource allocation and video quality adaptation by the cross-layer optimization method, in which fairness was also considered. In \cite{dubinvideo}, the fairness problem was addressed by applying a traffic shaping server.  

In \cite{liu2015fairness}, we studied the instability, unfairness, and underutilization problems for multiple DASH clients competing over a common bottleneck, where we proposed to probe the bandwidth by a logarithmic law based increase probing scheme and a conservative back-off based decrease probing scheme. Compared with \cite{li2014probe}, the probed bandwidth converges more quickly and tracks the time-varying bandwidth better. Besides, a dual-threshold buffer model was adopted to avoid buffer overflow/underflow. However, the method did not fully consider key factor that most impact visual quality, including buffer occupancy, video playback quality, video bit-rate switching frequency and amplitude. As an extension of \cite{liu2015fairness}, the work achieves a better trade-off among the instability, unfairness, and underutilization. Besides the LIMD based bandwidth probing scheme and dual threshold buffer model, a probability driven rate control logic is also proposed to achieve better visual quality. Moreover, it can break out the balance that the bandwidth is in perfect-subscription with unfair share of the bandwidth.

\begin{figure}[ht]
\centering
\includegraphics[width=0.46\textwidth]{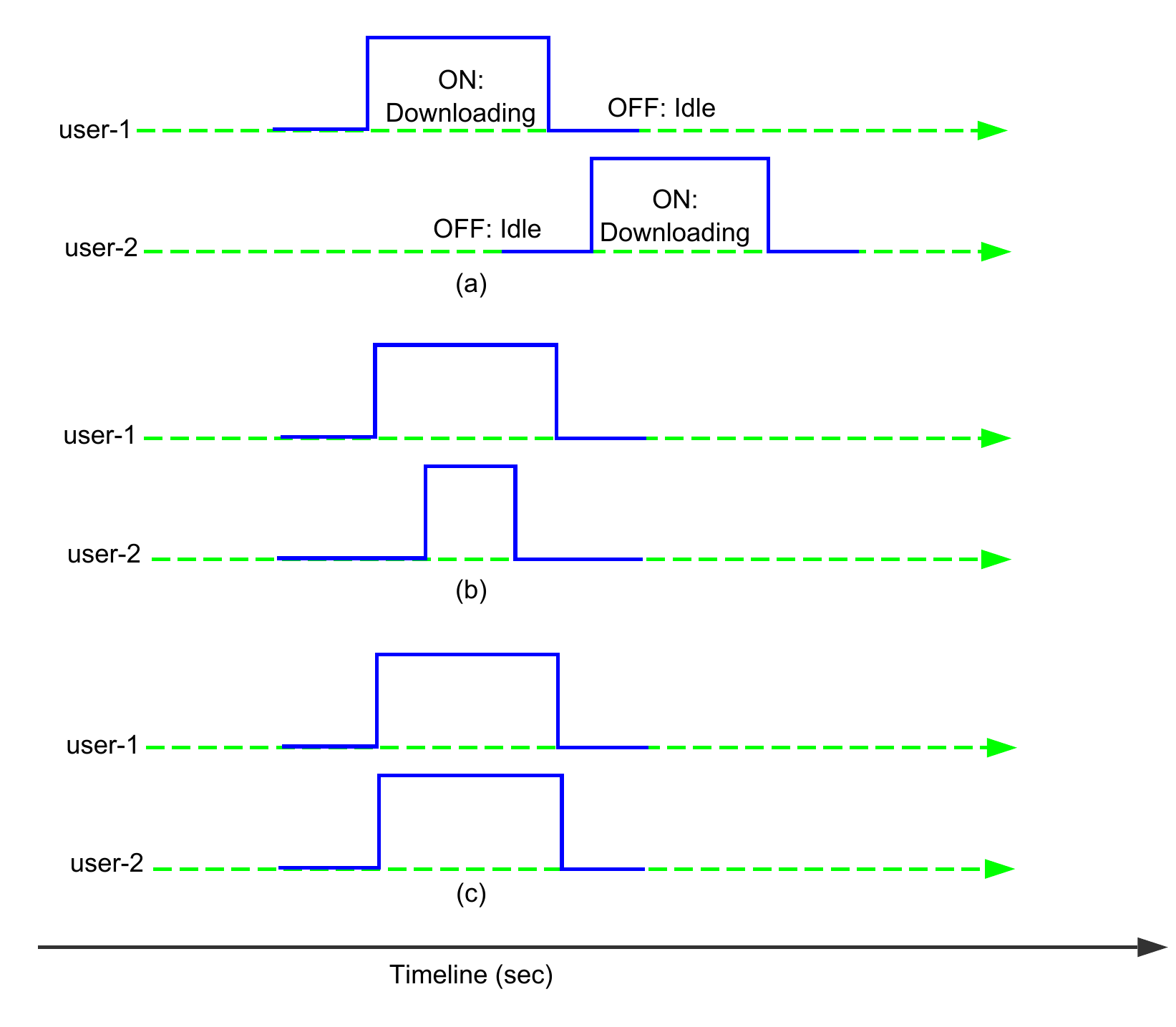}
\caption{Illustration two clients competing shared bandwidth under three scenarios of the ON-OFF periods.}
\label{fig:problem statement}
\end{figure}

\section{Problems with Rate Adaption for Multiple Clients over DASH}
\label{Sec:rate adaption}

\subsection{Problem Statement}
\label{subSec:problem statement}
To provide satisfactory performance for media streaming under the scenario of multiple users over DASH, the issues about stability, fairness, and  bandwidth utilization need to be jointly considered. To explain it clearly, Fig. \ref{fig:problem statement} illustrates the temporal overlap of the ON-OFF periods among two competing users as an example. Suppose the available bandwidth is $C$ and that a single active connection gets the whole available bandwidth, while two active connections share it fairly, i.e., each gets $C/2$. We further denote by $C_1$ and $C_2$ the throughput received by the two users, and $C_1 = C_2 = C/2$ ideally. Fig. \ref{fig:problem statement}(a) shows the case where there is no overlap of the ON periods for the two users, i.e., each user monopolizes the available bandwidth during its ON period such that we have $C_1 = C_2 = C$. When both users overestimate their fairly shared bandwidth, they may request new segments with higher bit-rates, thereby causing congestion. In the case of congestion, the users will find that their estimated bandwidth is less than their previous estimation, and thus will switch to a lower video bit-rate. This oscillation can repeat, leading to video playback {\textbf{instability}}. Fig. \ref{fig:problem statement}(b) shows the situation where the ON period of one user falls in the ON period of the other user. This may happen when the users request video segments with different video bit-rates. In this case, the throughput observed by the users is $C_1 > C/2$ and $C_2 = C/2$, i.e., user one overestimates the fair share of bandwidth. When only one user overestimates the fair share of bandwidth, the two users will converge to a stable but unfair equilibrium that the user, who overestimates the fair share of bandwidth, will request a higher video bit-rate, causing {\textbf{unfairness}} problem. At last, in Fig. \ref{fig:problem statement}(c) the ON periods of the two users are perfectly aligned, and both users observe that $C_1 = C_2 = C/2$. Though both users estimate the fair share of bandwidth correctly, it still may cause {\textbf{underutilization}} problem. This is mainly because the available bandwidth at the server side is discrete (quantized) and limited. For example, when the quantized available bandwidth cannot be $C/2$, then both users may request the maximum available video bit-rate which is smaller than $C/2$ to avoid playback interruptions (buffer underflow). In this case, OFF periods (sleeping mechanism) are adopted to prevent buffer overflow, and underutilization problem occurs.

For bandwidth consumption, three different scenarios are considered: perfect-subscription, over-subscription,  and under-subscription, denoting the situations that the total amount of bandwidth requested by the clients is equal to, larger than, and smaller than the available bandwidth, respectively.  Suppose TCP is in ideal behavior, i.e., perfectly equal sharing of the available bandwidth when the transfers overlap. The unfairness among clients mainly comes from the ON-OFF phenomenon in DASH \cite{akhshabi2012happens}. As demonstrated in \cite{li2014probe}, only when the bandwidth is in over-subscription, i.e., the congestion occurs, the bandwidths estimated by the clients are approximately equal, and fairness can be obtained. This is because when all the clients are continuously downloading without any OFF period, the bandwidth is equally shared by the DASH clients since all DASH clients are built on top of TCP which is fair in nature. However, when congestion occurs, the video buffer at the client side will be drained, causing playback freeze, which can severely degrade quality of experience. When the bandwidth is in under-subscription, the requested video bit-rate is smaller than the available bandwidth, and OFF periods are needed to suspend the transmission so as to avoid buffer overflow. In this case, instability, unfairness, and underutilization problems may happen as explained above.



Intuitively the objective of a rate adaption scheme is to control the bandwidth to stay in perfect-subscription, and all the clients request the same bit-rate as the fair-share bandwidth. However, it is very difficult, if not impossible, to achieve this objective since only when the bandwidth is in over-subscription, the fairness can be guaranteed for the DASH clients. On the other hand, even if the DASH clients can obtain the fair-share bandwidth, generally there is no available video bit-rate which exactly matches the fair-share bandwidth since the video sequences are only transcoded into several discrete bit-rates at the server side.

\begin{figure*}[ht]
\centering
\includegraphics[width=0.8\textwidth]{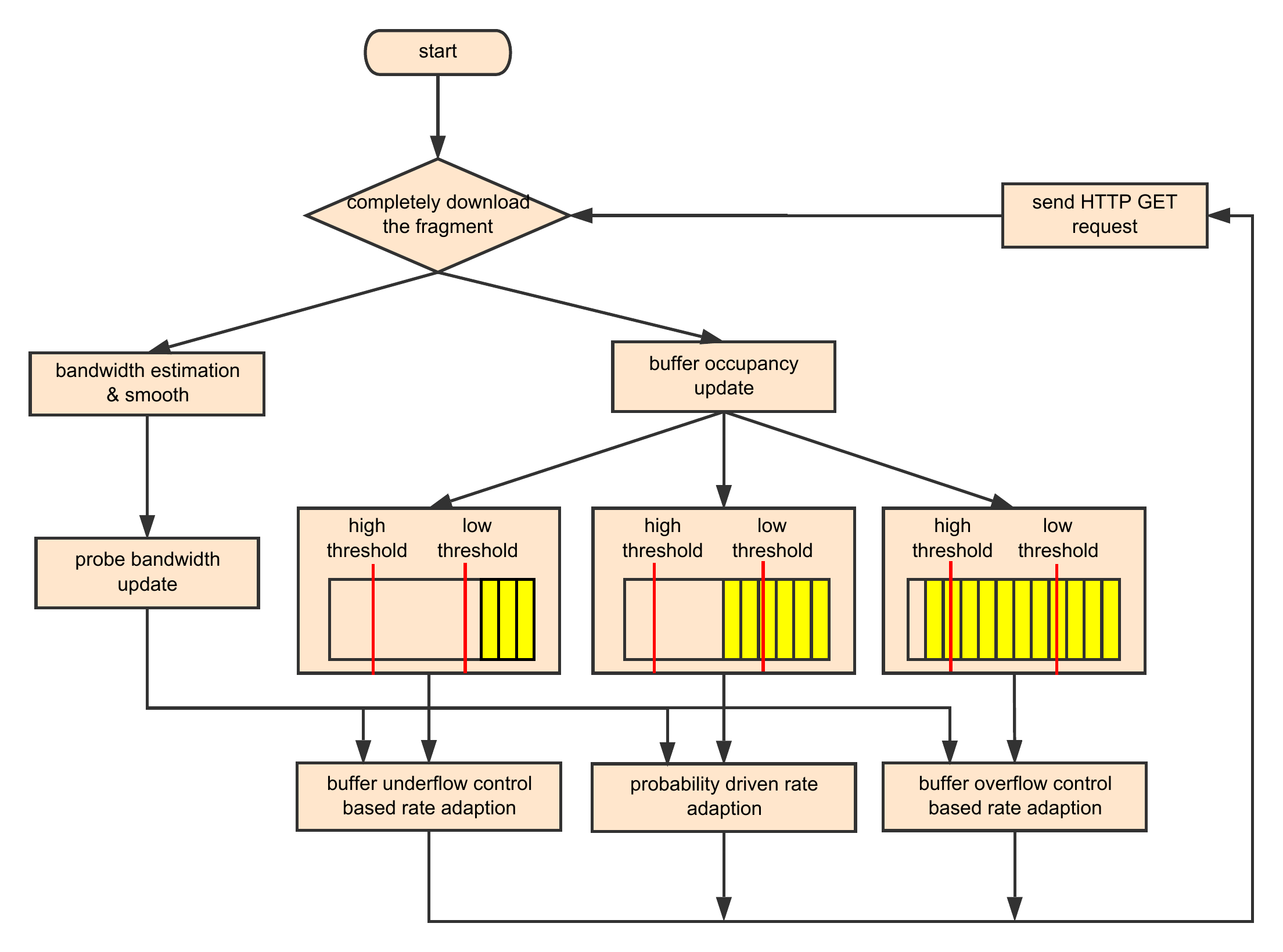}
\caption{Flowchart of the proposed rate adaption scheme for TFDASH.}
\label{fig:flowchart}
\end{figure*}

\subsection{Rate Adaption Approach Overview}
\label{subSec:rate adaption}
In this work, fairness is enforced on a long-term (not instantaneous) basis such that, for an certain moment, the clients may request different video bit-rates. While for relatively long time, the average video bit-rates for the clients are approximately equivalent. This can be achieved by alternately switching up and down the video bit-rate. However, from the viewpoint of quality of user experience, frequent bit-rate switching is annoying. Therefore, a dual-threshold buffered model is adopted, which was proposed and proven to be effective in our previous work \cite{r6}. To improve the visual quality, based on this model, we further propose a probability-driven rate switching logic, which takes into account several key factors that most impact visual quality, including buffer occupancy, video playback quality, video rate switching frequency and amplitude.

To ensure fairness, we focus on preventing the occurrence of OFF phenomenon, i.e., the bandwidth is either in perfect-subscription or in over-subscription. In the case of over-subscription, the fairness is guaranteed, whereas in the case of perfect-subscription, there is an unlimited number of bandwidth sharing modes. It is worth to notice that though it is easy to avoid OFF phenomenon by keeping on downloading without interruption, it is even impossible to distinguish if the bandwidth is in perfect-subscription or in over-subscription. Moreover, since the available bit-rate is limited and discrete, instead of using the estimated bandwidth, we propose an effective bandwidth probing scheme to guide the rate adaptation. As a result, the dynamic equilibrium state is probed for each client such that the client's video bit-rate fluctuates around the fair-share bandwidth, which reaches long-term fairness.

Fig. \ref{fig:flowchart} shows the flowchart of our proposed rate adaption approach. In DASH, the video bit-rate can be changed only when a segment is completely downloaded. Then, to determine the video bit-rate for the next segment to be downloaded, we first estimate the bandwidth and smooth it by a Kalman filter (Section \ref{subSec: bandwidth estimation}). Then, together with the estimated bandwidth, we update the probed bandwidth which will be used to guide the rate adaption (Section \ref{subsec: bandwidth probing}). According to the buffer occupancy, the video bit-rate is adapted under three scenarios. When the buffer occupancy is in between the two predefined thresholds, a probability driven rate adaption is applied (Section \ref{subSec:probability}); otherwise, the video bit-rate is selected to avoid buffer overflow and underflow (\ref{subSec:buffer control}).

The advantages of our proposed rate adaption scheme can be summarized as follows:

{\bf{Fairness}}: In our proposed rate adaptation approach, no OFF phenomenon happens unless the maximal available video bit-rate has been reached. Without OFF phenomenon, the bandwidth is either over-subscription or perfect-subscription. For the former, the fairness is guaranteed. While for the later, though there are unlimited number of bandwidth sharing modes in theory, depending on the starting time of downloads, instead of using the estimated bandwidth, we probe the video bit-rate by combining the probed bandwidth and probability driven rate adaption to break the balance, and bandwidth is transfered from perfect-subscription to over-subscription.

{\bf{Efficiency}}: Without OFF phenomenon, the available bandwidth can be shared by clients with high-efficiency, and the bandwidth utilization approaches to 100\%. The experiments also show that our proposed rate adaptation approach achieves high average video bit-rate and high bandwidth utilization efficiency.

{\bf{Stability}}: With the proposed dual-threshold, the effect of buffer occupancy oscillations and bandwidth variations on video bit-rate adaptation can be mitigated. Furthermore, during the rate adaptation process, the smoothness (i.e., the number of previous segments which have the same video bit-rate) is also taken into account.

{\bf{High visual quality}}: Besides the smoothness, our method also consider several additional key factors that most influence visual quality, including buffer occupancy, video playback quality, video rate switching frequency and amplitude. Besides, buffer overflow and underflow are effectively mitigated.

In the following sections, we give the details of the components in Fig. \ref{fig:flowchart} for the rate adaptation approach. Important symbols used in this paper are summarized in Table \ref{tab:symbol}.

\begin{table}[H]
\begin{center}
\caption{Major symbols used in the paper} \label{tab:symbol}
\begin{tabular}{ p{1.4cm}| p{6.6cm}}
  \hline
  \hline
   $\hat b_k$ & segment-bias bandwidth for segment $k$\\
  \hline
  ${{\mathord{\buildrel{\lower3pt\hbox{$\scriptscriptstyle\frown$}} \over b} }_k}$ & smoothed bandwidth for segment $k$ \\
  \hline
  ${\tilde b_k} $ & probed bandwidth for segment $k$ \\
  \hline
  $q(t)$ & buffer occupancy at time $t$ \\
  \hline
  $q_{low}$ & threshold of buffer occupancy to prevent buffer underflow \\
  \hline
  $q_{high}$ & threshold of buffer occupancy to prevent buffer overflow \\
  \hline
  $\tau$ & playback duration of each segment (in seconds) \\
  \hline
  $t^{s}_k$ & time instant when start to download segment $k$ \\
  \hline
  $t^{e}_k$ & time instant when segment $k$ is downloaded completely\\
  \hline
  $V_i$ & $i$-th available video bit-rate with $i = 1, 2, ..., L$ \\
  \hline
  $v_k$ & video bit-rate for segment $k$\\
 \hline
 \hline
\end{tabular}
\end{center}
\end{table}

\section{Receiving Bandwidth Estimation and Probing}
\label{Sec:bandwidth}

\subsection{Smooth TCP Receiving Bandwidth Estimation}
\label{subSec: bandwidth estimation}
Assume a video clip is encoded into  $L$ different versions, with different playback video bit-rates $V_1<V_2< ... <V_L$. All versions of the video are partitioned into equal-length segments, each of which consuming the same playback time of $\tau$. For each client, the streaming process is divided into sequential segment downloading steps $k = 1, 2, 3, ...$. Without loss of generality, suppose a client starts downloading segment $k$ at time instant $t^{s}_k$ and the segment is downloaded completely at $t^{e}_k$. and the video bit-rate for segment $k$ is $v_k$. Then, the smoothed segment-bias bandwidth, $\hat b_k$, can be estimated as \cite{tmm-16}:
\begin{equation}
{\hat b_k} = \frac{{{v_k}\tau }}{{t_k^e - t_k^s}}
\end{equation}

Furthermore, to eliminate the noise and interference during the receiving bandwidth estimation, Kalman filter \cite{friedland1980steady,dolinsky2012kalman} is adopted. Then, we have
\begin{equation}
\label{eq:estimated bandwidth}
{{\mathord{\buildrel{\lower3pt\hbox{$\scriptscriptstyle\frown$}}
\over b} }_k} = w {{\hat b}_{k-1}} + \left( {1 - w } \right){{\mathord{\buildrel{\lower3pt\hbox{$\scriptscriptstyle\frown$}}
\over b} }_{k - 1}}
\end{equation}
where ${{\mathord{\buildrel{\lower3pt\hbox{$\scriptscriptstyle\frown$}} \over b} }_k}$ is the amended bandwidth by Kalman filter, and is $w$ a coefficient with $0 < w < 1$. Eq. (\ref{eq:estimated bandwidth}) shows that the amended bandwidth for downloading segment $k$, ${{\mathord{\buildrel{\lower3pt\hbox{$\scriptscriptstyle\frown$}} \over b} }_k}$, is adjusted by $w$ based on the previous amended bandwidth  ${{\mathord{\buildrel{\lower3pt\hbox{$\scriptscriptstyle\frown$}} \over b} }_{k-1}}$ and the currently measured bandwidth ${{\hat b}_{k-1}}$. If $w$ is small, the previous amended bandwidth may play a more important role in predicting the bandwidth, As $w$ gets larger, the amended bandwidth  is closer to the current measured value. Thus, if the variation of bandwidth is large, we should decrease $w$, and if the variation of bandwidth is small, we can increase $w$ to more accurately reflect the change of network condition. Therefore, the value of coefficient $w$ is given as follows:
\begin{equation}
w = \frac{1}{{1 + {e^{u - {u_0}}}}}
\end{equation}
where $u_0$ is a constant, and $u$ is the normalized difference between the measured value and amended value that
\begin{equation}
u = \frac{{\left| {{{\hat b}_{k - 1}} - {{\mathord{\buildrel{\lower3pt\hbox{$\scriptscriptstyle\frown$}}
\over b} }_{k - 1}}} \right|}}{{{{\hat b}_{k - 1}}}}
\end{equation}

\subsection{LIMD-based Bandwidth Probing}
\label{subsec: bandwidth probing}
Due to the ON-OFF phenomenon in DASH, the bandwidth estimated by a client is discrepant and cannot be used directly for rate adaptation \cite{akhshabi2012happens}. Moreover, as demonstrated in \ref{subSec:problem statement}, only when the bandwidth is oversubscribed, i.e., the congestion occurs, the bandwidth estimated by a client is equal to the fair-share bandwidth (all the clients see nearly the same available bandwidth). On the other hand, when congestion occurs, the requested video bit-rate cannot be supported by the bandwidth, and playback freeze may happen. Thus, how to obtain the fair-share bandwidth without congestion is critical for improving the rate adaptation performance for DASH.

In this section, we propose a Logarithmic Increase Multiplicative Decrease (LIMD) based bandwidth probing scheme, which includes a logarithmic increase phase and a multiplicative decrease phase, to  make the probed bandwidth quickly converge to the fair-share bandwidth. The intuition behind this method is that the estimated bandwidth in \ref{subSec: bandwidth estimation} by a client is always the upper bound of the fair-share bandwidth due to the off intervals \cite{liu2015fairness,akhshabi2012happens}. Thus, during the increase probing phase, the probed bandwidth will continuously increase until it exceeds the estimated bandwidth in (\ref{eq:estimated bandwidth}). Then, when the probed bandwidth is higher than the estimated bandwidth, the client will switch to the decrease probing phase to avoid congestion. This is similar to the Additive Increase Multiplicative Decrease (AIMD) based congestion control of TCP, where the transport rate continuously increases until a packet loss happens.

However, there are many problems with AIMD-based scheme in video streaming, such as its slow start and frequent fluctuations. Therefore,  the granularity of probing has to be selected carefully \cite{liu2015fairness,li2014probe}. Coarse granularity makes the probed bandwidth converge quickly, but it may over-probe and lead to congestion. In contrast, probing of too fine granularity takes a long time to converge and usually cannot track the time-varying bandwidth well.

To address the above issues, an LIMD based probing scheme is proposed to increase the probed bandwidth since when the gap between the probed bandwidth and the fair-share bandwidth (we cannot obtain the fair-share bandwidth in practice, and instead its upper bound, i.e., the estimated bandwidth is used for approximation) is large, a coarse granularity probing scheme should be employed to increase the probed bandwidth quickly. On the other hand, when the gap is small, a fine granularity probing scheme should be adopted to avoid over-probing. Moreover, when the probed bandwidth is higher than the fair-share bandwidth, congestions may occur, thereby leading to playback freezing. In this case, a conservative back-off scheme is designed to decrease the probed bandwidth guaranteeing that no congestion happens.

Specifically, we denote ${\tilde b_k} $ the probed bandwidth which will be used for guiding the video bit-rate switching for  segment $k$, and it is initialized to zero that ${\tilde b_0} = 0$. Whenever a segment is completely downloaded, the estimated bandwidth is update according to (\ref{eq:estimated bandwidth}). Then, the probed bandwidth is updated as follows:
\begin{equation}
{\tilde b_k}  = \left\{ \begin{array}{l l}
\tilde b_{k-1} + \text{max} (\frac{{{\mathord{\buildrel{\lower3pt\hbox{$\scriptscriptstyle\frown$}} \over b} }_{k-1}}  - \tilde b_{k-1}}{2},\bigtriangleup ),   \\ \quad \quad \quad \quad \text{if}\; \tilde b_{k-1} < {{\mathord{\buildrel{\lower3pt\hbox{$\scriptscriptstyle\frown$}} \over b} }_{k-1}} \\\\
\tilde b_{k-1} + \alpha ({{\mathord{\buildrel{\lower3pt\hbox{$\scriptscriptstyle\frown$}} \over b} }_{k-1}}  - \tilde b_{k-1}),  \\ \quad \quad \quad \quad  \text{if}\; \tilde b_{k-1} \ge {{\mathord{\buildrel{\lower3pt\hbox{$\scriptscriptstyle\frown$}} \over b} }_{k-1}}
\end{array}\right.
\label{eq:probe bandwidth}
\end{equation}
where $\bigtriangleup$ is a constant to avoid slow convergence, and  $\alpha$ is a positive constant satisfying that $\alpha > 1$. By (\ref{eq:probe bandwidth}), the probed bandwidth will quickly approach to the estimated bandwidth in the logarithmic law at first. However, when the gap between the $\tilde b_{k-1}$ and ${{\mathord{\buildrel{\lower3pt\hbox{$\scriptscriptstyle\frown$}} \over b} }_{k-1}}$ is small, i.e., $\frac{{{\mathord{\buildrel{\lower3pt\hbox{$\scriptscriptstyle\frown$}} \over b} }_{k-1}}  - \tilde b_{k-1}}{2} < \bigtriangleup$, the probed bandwidth would be additively increased by $\bigtriangleup$ rather than by logarithmic law. On the other hand, when the probed bandwidth exceeds the estimated, a conservative back-off scheme is adopted to control $\tilde b_{k-1}$ to be not higher than ${{\mathord{\buildrel{\lower3pt\hbox{$\scriptscriptstyle\frown$}} \over b} }_{k-1}}$.

\section{Rate Adaptation Logic}
\label{Sec:rate adaption}

\subsection{Buffered Video Time Model}
\label{subSec:buffered video time}
To maintain continuous playback, a video streaming client normally contains a video buffer to absorb temporary mismatch between the video downloading rate and video playback rate. In conventional single-version video streaming, the buffered video playback time can be easily measured by dividing the buffered video size by the average video playback rate. In DASH, however, different video versions have different video playback rates. Since a video buffer contains segments from different versions, there is no longer a direct mapping between the buffered video size and the buffered video time. To tackle the problem, we use the buffered video time to measure the length of video playback buffer.

The buffered video time process, represented as $q(t)$, can be modeled as a queue with a constant service rate of unity, i.e., in each second, a piece of video with length of one second of playback time is dequeued from the buffer and then played. The enqueue process is driven by the video download rate and the downloaded video version. We adapt the video bit-rate when a segment has been downloaded completely. Without loss of generality, suppose a client starts downloading segment $k$ at time instant $t^{s}_k$ and the segment is downloaded completely at $t^{e}_k$.  Then, we have
\begin{equation}
t_{k+1}^e  - t_{k+1}^s  = \frac{{v_{ k+1}}}{{ {\mathord{\buildrel{\lower3pt\hbox{$\scriptscriptstyle\frown$}}
\over b} }_k }} \tau
\end{equation}
and ${\mathord{\buildrel{\lower3pt\hbox{$\scriptscriptstyle\frown$}} \over b} }_k$ is the estimated bandwidth given in (\ref{eq:estimated bandwidth})). Then the buffered video time evolution becomes
\begin{equation}
\label{eq:buffer evolution}
q\left( {t_{k+1}^e } \right) = q\left( {t_{k+1}^s } \right) + \tau - \frac{{v_{k+1}}}{{ {\mathord{\buildrel{\lower3pt\hbox{$\scriptscriptstyle\frown$}}
\over b} }_k }} \tau
\end{equation}
where the second term of (\ref{eq:buffer evolution}) is the added video time upon the completion of the downloading of segment $k+1$, and the third term reflects the fact that the buffered video time is consumed linearly at a rate of unity during the downloading process. However, if the bandwidth is too high, a sleep mechanism is needed to postpone the segment request so as to avoid buffer overflow \cite{r6}. Assume the sleeping time is $\tau _s$, then (\ref{eq:buffer evolution}) is rewritten as
\begin{equation}
q\left( {t_{k + 1}^e } \right) = q\left( {t_{k+1}^s } \right)  + \tau - \frac{{v_{k+1} }}{{{\mathord{\buildrel{\lower3pt\hbox{$\scriptscriptstyle\frown$}}
\over b} }_k }} \tau  - \tau _s
\label{eq:nextbuffer2}
\end{equation}

From the control system point of view, there is a fundamental conflict between maintaining stable video bit-rate and maintaining stable buffer occupancy, due to the unavoidable network bandwidth variations. Nevertheless, from the end user point of view, video bit-rate fluctuations are much more perceivable than buffer occupancy oscillations. The recent work in \cite{r14} reported that switching back-and-forth between different bit-rates will significantly degrade users¡¯ viewing experience, whereas buffer size variations do not have direct impact on video streaming quality as long as the video buffer does not deplete. Thus, a dual-threshold buffer occupancy model is adopted to mitigate the effect of buffer occupancy oscillations on video bit-rate adaptation. The dual-threshold buffer occupancy model has two predefined thresholds, $q_{high}$ and $q_{low}$.

When the buffer occupancy is in between $q_{high}$ and $q_{low}$, the video bit-rate is adapted in a probabilistic manner by taking into account several key factors that have critical impact on visual quality, including buffer occupancy, video playback quality, video rate switching frequency and amplitude. Otherwise, video bit-rate is switched to avoid buffer overflow/underflow. The details are given in the next subsections.

\subsection{Probability-Driven Rate Control}
\label{subSec:probability}
\begin{figure*}
\begin{equation}
\label{eq:probabilty}
\begin{aligned}
&P\left( {{v_k}\left| {{q_{k - 1}},{v_{k - 1}},{n_{k - 1}}} \right.} \right)\\
 = & \underbrace {\left( {\frac{{1 + {\mathop{\rm sgn}} \left( {{v_k} - {v_{k - 1}}} \right)}}{2}f\left( {{q_{k - 1}}} \right) + \frac{{1 - {\mathop{\rm sgn}} \left( {{v_k} - {v_{k - 1}}} \right)}}{2}\left( {1 - f\left( {{q_{k - 1}}} \right)} \right)} \right)}_{{C_1}}\\
 * &\underbrace {\frac{{\ln \left( {{v_k} - {v_{\min }} + \varepsilon } \right)}}{{\ln \left( {{v_{\max }} - {v_{\min }} + \varepsilon } \right)}}}_{{C_2}} * \underbrace {\left( {1 - \frac{{\ln \left( {\left| {{v_k} - {v_{k - 1}}} \right| + \varepsilon } \right)}}{{\ln \left( {{v_{\max }} - {v_{\min }} + \varepsilon } \right)}}} \right)}_{{C_3}} * \underbrace {f\left( {{n_{k - 1}}} \right)}_{{C_4}}
\end{aligned}
\end{equation}
\end{figure*}

When the buffer occupancy is in between the two thresholds, i.e., $q_{\min} \le q_{k-1} \le q_{\max}$, the continuous video playback can be guaranteed with high confidence. Then, we prefer to switch the video bit-rate to a more suitable level. Considering the fairness of all the competing clients, the video bit-rate is adapted in a probabilistic manner.

The probability is given by (\ref{eq:probabilty}), which is determined by several factors, including the bit-rate of last segment $v_{k-1}$, the current buffer occupancy $q_{k-1}$, the number of consecutive segments which have the same video bit-rate with segment $k-1$, and the target video bit-rate for the segment to be requested (i.e., bit-rate $v_k$ for segment $k$). The probability is designed by comprehensively considering the effects of key factors on quality of user experience, including the risk of buffer overflow and underflow, the video bit-rate switching frequency and amplitude, and the video quality. The probability in (\ref{eq:probabilty}) consists of four items. Now, we give the details of the motivation and explanation for each item as follows.

\emph{\textbf{Effect of buffer occupancy on video bit-rate switching decision}}, $C_1$. It is the sum of two sub-items, respectively denoting the switch-up and switch-down of video bit-rate, compared with the current video bit-rate. Each sub-item can be further roughly classified into three cases: $i$) the buffer occupancy is lower than the reference threshold $q_{ref}$, $ii$) the buffer occupancy is equal to $q_{ref}$, and $iii$) the buffer occupancy is higher than $q_{ref}$.

Now, we take the first sub-item, i.e., switch-up of video bit-rate as an example to analyze the above three cases. When the buffer occupancy is lower than $q_{ref}$, denoting the risk of buffer underflow is higher than that of buffer overflow in the bias of buffer occupancy. Thus, switch-up of video bit-rate is inadvisable. Besides, the larger of the gap between the buffer occupancy and $q_{ref}$, the higher of the risk of buffer underflow, and also the smaller of the probability of video bit-rate switch-up, and vice versa. At last, when the buffer occupancy is equal to $q_{ref}$, the risk of buffer underflow is the same as that of buffer overflow, and the probabilities of switch-up and switch-down of video bit-rate should be equal. Moreover, the probability of switch-up of video bit-rate should not be linearly proportional to the buffer occupancy. Instead, the probability should decrease quickly when the buffer occupancy goes below $q_{ref}$ to avoid buffer underflow, and increase quickly when the buffer occupancy goes above $q_{ref}$ to avoid buffer overflow.  Therefore, the following Sigmoid function is applied since it meets the above requirements:
\begin{equation}
\label{eq:sigmoid}
f\left( x \right) = \left\{ {\begin{array}{*{20}{l}}
0&{\rm{if}}\;\;x < {x_{\min }}\\
1&{\rm{if}}\;\;x > {x_{\max }}\\
{\frac{1}{{1 + {e^{{x_0} - x}}}}}&{\rm{else}}
\end{array}} \right.
\end{equation}
where in $C_1$, $f(q) = f(x)$ by setting $x_{\min} = q_{\min}, x_{\max} = q_{\max}, x_0 = q_{ref}$, and ${\rm{sgn}}(x)$ is given in (\ref{eq:sgn}):
\begin{equation}
\label{eq:sgn}
{\mathop{\rm sgn}} \left( x \right) = \left\{ {\begin{array}{*{20}{l}}
1&{\rm{if}}\;\;x > 0\\
0&{\rm{if}}\;\;x = 0\\
{ - 1}&{\rm{if}\;\;x < 0}
\end{array}} \right.
\end{equation}
denoting if the video bit-rate is switched up or down, or remains unchanged.

The second sub-item in $C_1$ (i.e., switch-down of video bit-rate) has opposite characteristics with the case of video bit-rate switch-up, and the probability is given by $1-f(x)$.

\emph{\textbf{Effect of video quality on video bit-rate switching decision}}, $C_2$ . Generally, the higher the video bit-rate, the better the quality of experience. Thus, only in the bias of video bit-rate, a high video bit-rate is preferred during the rate switching process. On the other hand, as shown in \cite{r26, chao2016mdash}, user experience follows the logarithmic law, and QoE function can be modeled in a logarithmic form for applications of file downloading and web browsing. As such, we use the logarithmic function of the video bit-rate to present the effect of video bit-rate on the probability of video bit-rate switching. In $C_2$, $\varepsilon$  is a small positive number with $\varepsilon \ge 1$  to ensure that the probability is nonnegative, typically we set $\varepsilon = 1$, and the term of $\ln \left( {{v_{\max }} - {v_{\min }} + \varepsilon } \right)$ is used to normalize the probability to range $[0, 1]$.

\emph{\textbf{Effect of switching amplitude, i.e., the gap of the video bit-rate between two consecutive segments, on bit-rate switching decision}}, $C_3$. It was shown in \cite{r14} that gradually switching the video bit-rate is more preferred. Therefore, the probability of video bit-rate switching should be decreased as the gap of the video bit-rate increases. Also, the probability is modeled as the logarithmic function of the difference of the video bit-rate between two consecutive segments as shown in $C_3$, and parameter $\varepsilon \ge 1$.

\emph{\textbf{Effect of switching frequency (i.e., the smoothness of video bit-rate) on the bit-rate switching decision}}, $C_4$. Because frequent bit-rate switching is annoying \cite{r4}, the effect of bit-rate switching on visual quality decreases as the number of previous segments with the same video bit-rate increases. This is mainly because human is sensitive to frequent short-term video bit-rate fluctuations rather than long-term video bit-rate switchings. We use the number, which denotes the number of previous segments with the same video bit-rate, to present the smoothness of video bit-rate. The lager the number is, the higher the smoothness will be, and the less impact the video bit-rate switching will make on visual quality. Therefore, the probability is an increasing function of the number, and the Sigmoid function in (\ref{eq:sigmoid}) is adopted so that $f(n_k) = f(x)$ with $x_{\min} = n_{\min}, x_{\max} = n_{\max}$ and $x_0 = n_0$. Typically, we set $n_{\max} = N = 15$ which is a length of 30s of video content with segment length $\tau = 2$s, $n_{\min} = 1$ which is the minimum value of $n_k$, i.e., the latest two consecutive segments have different video bit-rates, and $n_0 = \frac{2*n_{\max}}{3}$ as smooth video bit-rate is more preferred when the buffer occupancy is in between $q_{\min}$ and $q_{\max}$.

After deriving the probabilities in (\ref{eq:probabilty}) for all candidate video bit-rates, the video bit-rate is switched to the target one for segment $k$ according to the probabilities.


\subsection{Buffer Overflow and Underflow Control}
\label{subSec:buffer control}
When the buffer occupancy is higher than $q_{high}$ or lower than $q_{low}$, buffer overflow or underflow should be avoided. To this end, when $q(t_k^s) > q_{high}$ or $q(t_k^s) < q_{low}$, an appropriate video bit-rate should be selected. 

Specifically, when buffer occupancy is low (lower than the threshold $q_{low}$), we will select the maximal available video bitrate which is lower than the estimated bandwidth. On the other hand, when when buffer occupancy is high (higher than the threshold $q_{high}$), the minimal available video bitrate which is higher than the estimated bandwidth is selected. Therefore, the video bit-rate for segment $k$ is given as:
\begin{equation}
{v_k} = \left\{ {\begin{array}{*{20}{l}}
{\mathop {\max }\limits_{1 \le i \le L} \left ( {{V_i}\left| {{V_i \le \hat b_k}} \right.} \right )}&{{\rm{if}}\;\;q(t_k^s) < q_{low}}\\
{\mathop {\min }\limits_{1 \le i \le L} \left ( {{V_i}\left| {{V_i \ge \hat b_k}} \right.} \right )}&{{\rm{if}}\;\;q(t_k^s) > q_{high}}
\end{array}} \right.
\end{equation}
and when $q_{low} \le q(t_k^s) \le q_{high}$, the video bit-rate is selected according to Section \ref{subSec:probability}.

\section{Experiments}
\label{Sec:experiment}
In this section, we evaluate our rate adaption algorithms in terms of efficiency, fairness, and stability under the scenario that multiple clients compete over a bottleneck link.

\begin{figure*}[ht]
\centering
\subfigure[Bit-rate performance of FESTIVE clients]{
\label{fig:rate_festive}
\includegraphics[width=0.46\textwidth]{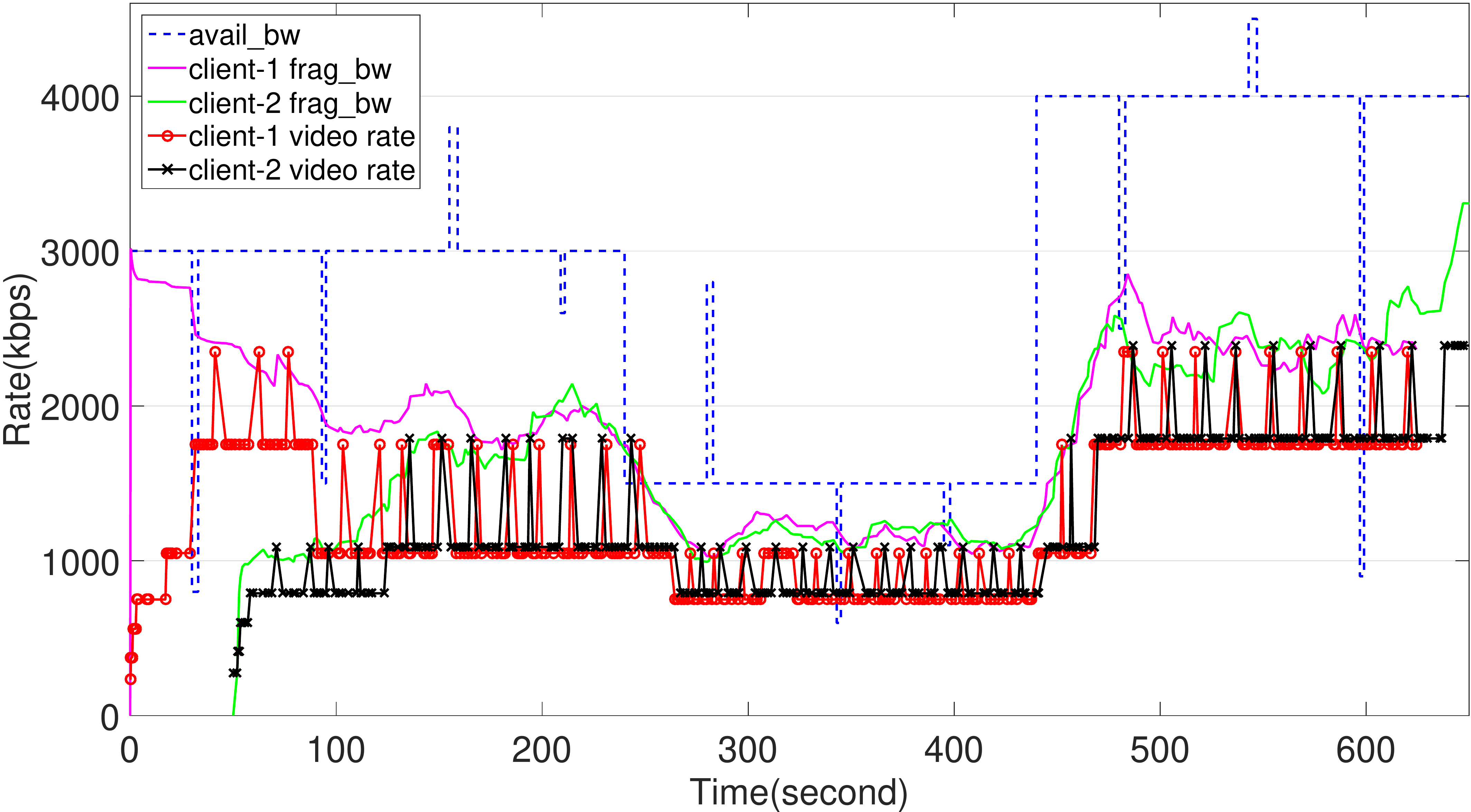}}
\hspace{0cm}
\subfigure[Buffer occupancy of FESTIVE clients]{
\label{fig:buffer_festive}
\includegraphics[width=0.46\textwidth]{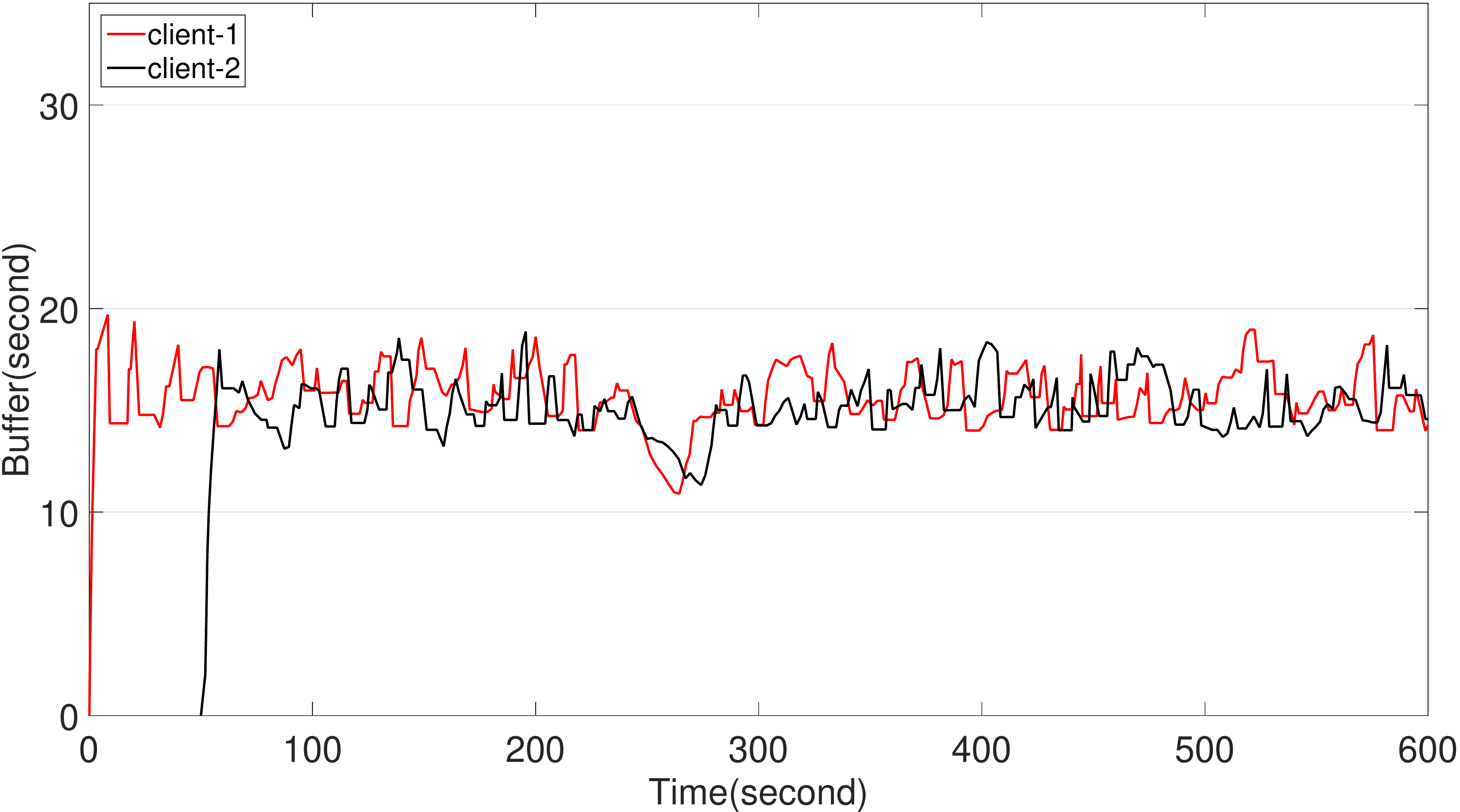}}
\hspace{0cm}
\subfigure[Bit-rate performance of PANDA clients]{
\label{fig:rate_panda}
\includegraphics[width=0.46\textwidth]{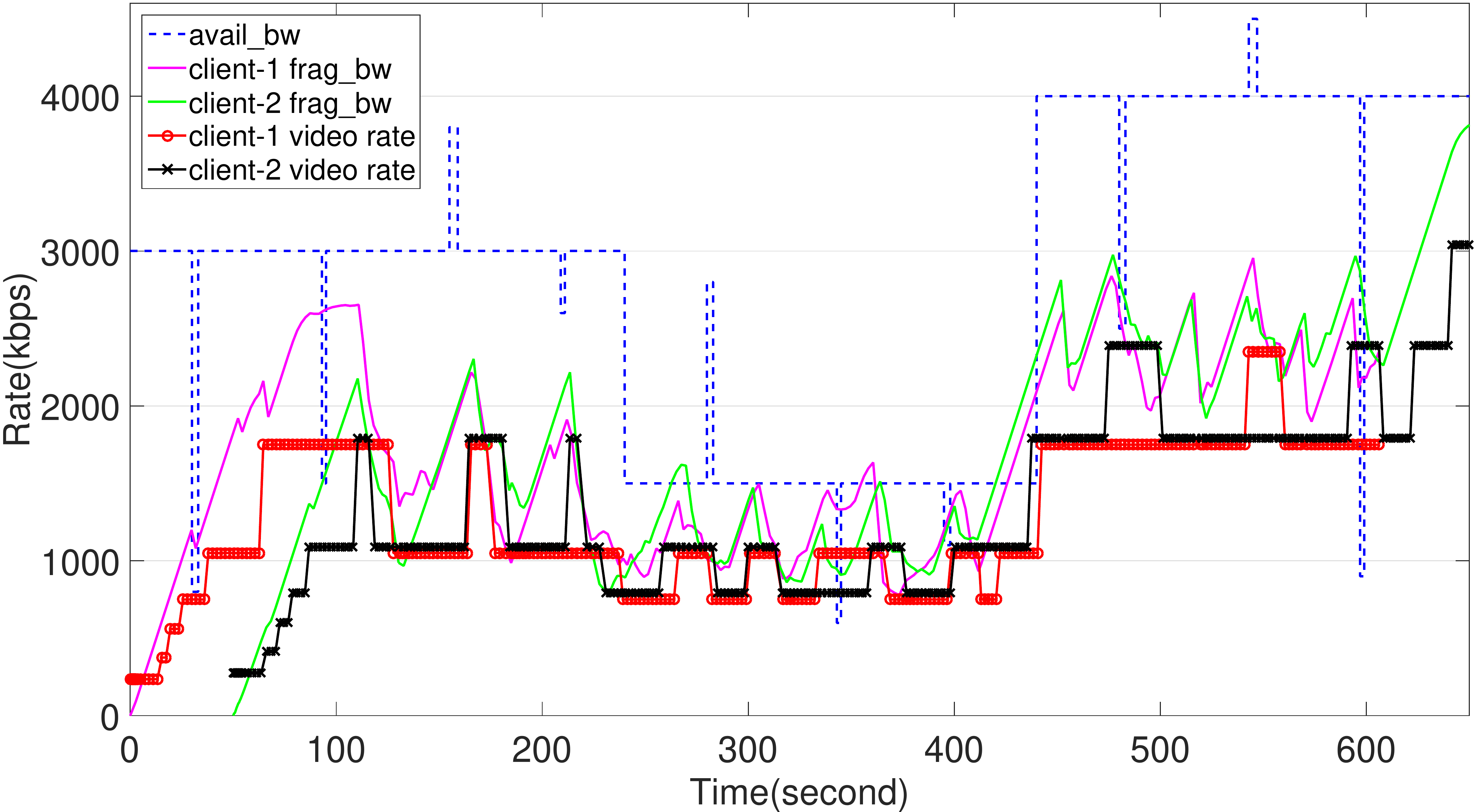}}
\hspace{0cm}
\subfigure[Buffer occupancy of PANDA clients]{
\label{fig:buffer_panda}
\includegraphics[width=0.46\textwidth]{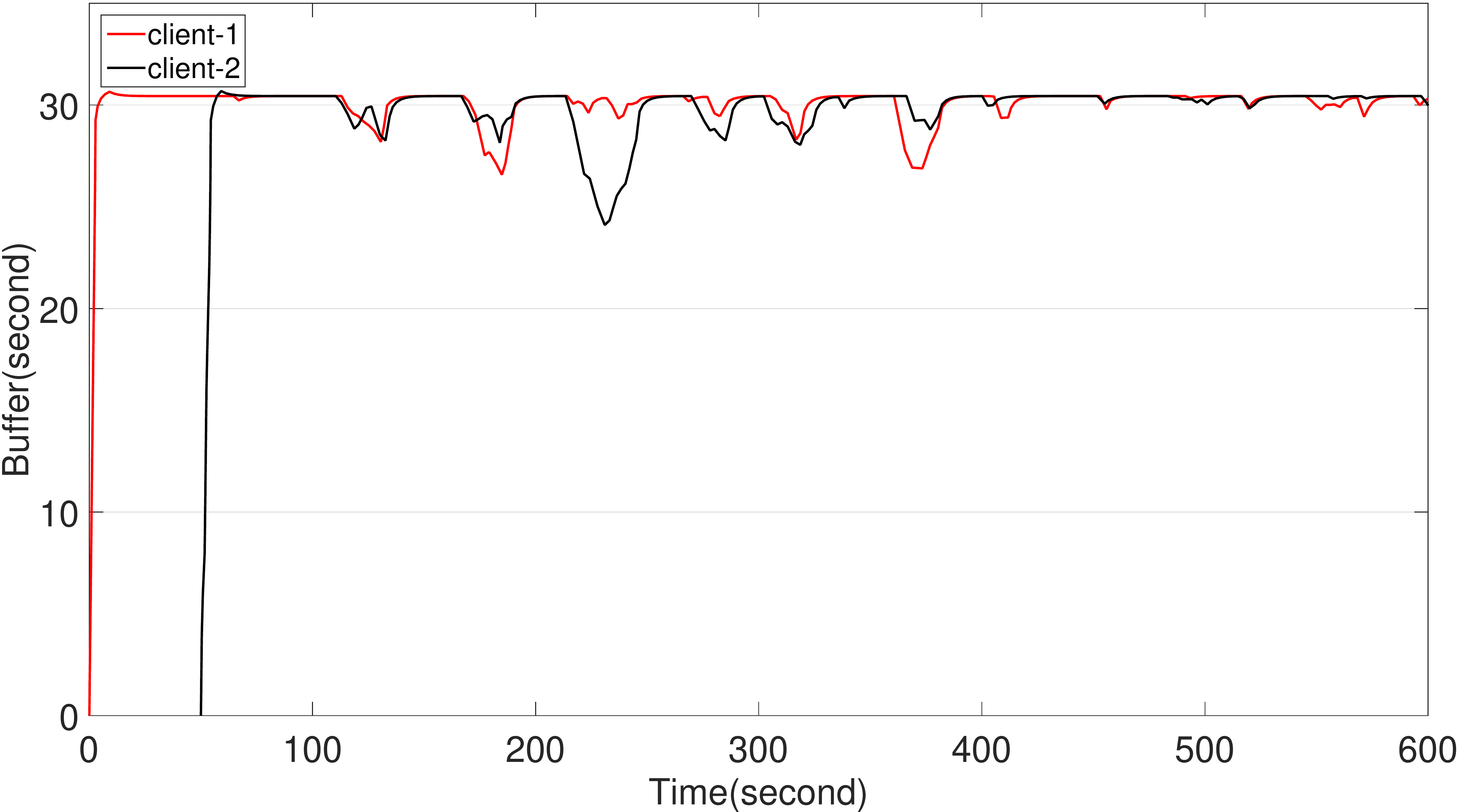}}
\hspace{0cm}
\subfigure[Bit-rate performance of TFDASH clients]{
\label{fig:rate_our}
\includegraphics[width=0.46\textwidth]{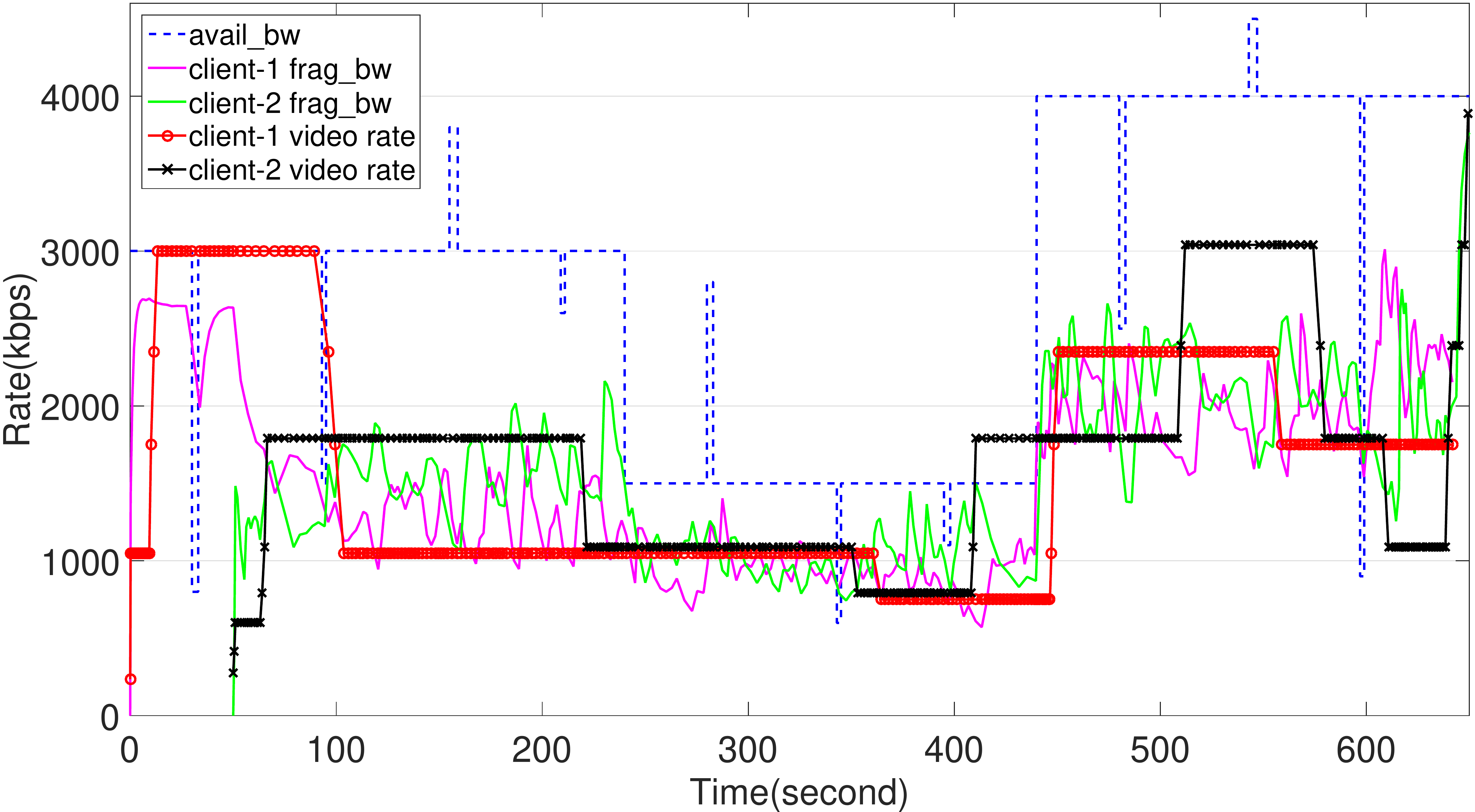}}
\hspace{0cm}
\subfigure[Buffer occupancy of TFDASH clients]{
\label{fig:buffer_our}
\includegraphics[width=0.46\textwidth]{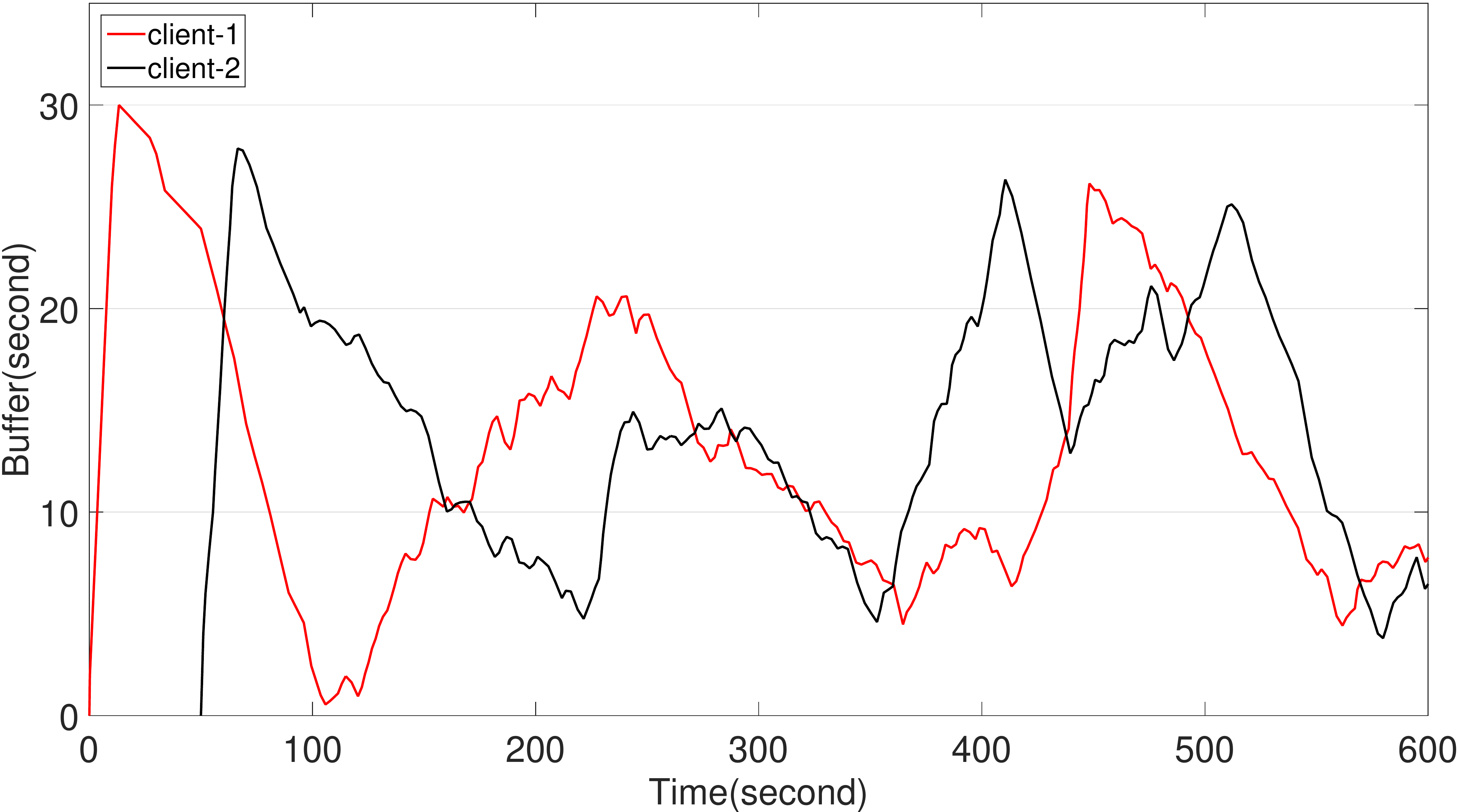}}
\caption{Video bitrate and buffer size evolution results under the same bandwidth variations described in Section \ref{subsection:bitrate switch performance}.}
\hspace{0cm}
\label{fig:rate-buffer}
\end{figure*}

\subsection{Experiments Setup}
We implement our TFDASH system in Linux/UNIX-based platforms. Our testbed consists of one web server handling media delivery, one router, and $n (n \ge 2)$ DASH clients. The server and clients run the standard Ubuntu of version 12.04.1. The server is installed with the Apache HTTP server of version 2.4.1, and Dummynet is used in the server to control the upload bandwidth \cite{r36}. In the testbed, the bottleneck is the bandwidth between the server and the router. In our experiments, for the content, we adopted the Big Buck Bunny sequence which is also part of the DASH dataset \cite{contentrul}.  The video sequence is pre-transcoded into 11 different video bitrates with three resolutions that the resolutions is 480p for bitrate of {235, 375, 560} kbps, 720p for {750, 1050, 1750, 2350} kbps, and 1080p for {3000, 3850, 4300, 5800} kbps.

Due to the complex network characteristics, it is hard to find the optimal values of the two thresholds $q_{high}$ and $q_{low}$ \cite{r6}. In our experiments, we set $q_{low} = 5$s since this start-up delay can be tolerated by most existing streaming systems, and $q_{high} = 25$s which is reasonable value for current existing devices, including mobile phones, to buffer such a length of media data. The maximum buffer size is set to 30s for all schemes in this work. However, generally a better performance can be obtained with a larger buffer size. This is because with a larger buffer size, the bandwidth variations can be better absorbed by the buffered video and a smoother video rate is guaranteed. Besides, we can also set relatively large $q_{low}$ to maintain a higher buffer occupancy so as to ensure continuous video playback. The effects of these two parameters on system performance can be found in our previous work \cite{r6}. For the other parameters used in the experiments, there default values are $\alpha = 1.25$, $\Delta = 32$kbps, and $u_0 = 0.5$.

For performance comparison, besides our TFDASH method, we also implement some popular DASH clients, including FESTIVE \cite{jiang2014improving} and PANDA \cite{li2014probe}. FESTIVE was proposed to balance the performance of efficiency, fairness, and stability for DASH clients by subtracting a random buffered video time from the target buffer occupancy for each segment request. In contrast, PANDA particularly considers the scenario when multiple clients are competing over a bottleneck link. It adopts the additive-increase multiplicative-decrease (AIMD) scheme to probe the available bandwidth, which is similar to the TCP congestion control scheme, and then adjust the video bit-rate according to the probed bandwidth.

\subsection{Video Bit-Rate Switching Performance}
\label{subsection:bitrate switch performance}
In this experiments, the bandwidth between the server and the router is controlled by Dummynet such that, during the first 230s, the available bandwidth (avail\_bw) is set to be 3000 kbps, and then it is limited to be 1500 kbps from 230s to 440s. At last, the avail\_bw is switched to 4000 kbps until the experiment ends. While during the first phase, i.e., from 0s to 230s, four short-term bandwidth variations are added, including both positive and negative spikes.  And for the second phrase (from 230s to 440s) and third phrase (from 440s to 650s), several positive and negative bandwidth spikes are added. Besides, two clients are competing for the avail\_bw (for the case of more clients, the performance is shown in the next subsection), the first client begins to download the video segments at the beginning of the experiment until all the video segments have been downloaded completely, and then it leaves the system. The second client joints the system at 50s, and leaves when it has completely downloaded all video segments.

The results of all DASH clients are shown in Fig. \ref{fig:rate-buffer}. First, the performances on rate and buffer occupancy with FESTIVE are shown in Fig. \ref{fig:rate_festive} and Fig. \ref{fig:buffer_festive}, respectively. Since FESTIVE does not apply the probing scheme, and instead it directly utilize the estimated bandwidth for rate adaptation, we plot the estimated bandwidth (i.e., segment-level bandwidth, frag\_bw for short, which is obtained as the segment size divided by the time to download the segment \cite{r25}) for each client. Due to the ON-OFF operations, the estimated bandwidth is generally higher than the fairly-shared bandwidth. Fig. \ref{fig:rate_festive} shows the sum of the two clients' estimated bandwidths is generally higher than the available bandwidth, while both clients have close estimated bandwidth. Thus, the clients usually over-estimate the fairly-shared bandwidth and then request for video segments with too high video bit-rate, thereby causing congestions. When this happens, the clients will find that their estimated bandwidth is less than their previous estimation, and thus switch to a lower video bit-rate. This oscillation can repeat, thereby causing video playback instability as explained in Fig. \ref{fig:problem statement}. Fig. \ref{fig:rate_festive} also demonstrates this the video bit-rates of clients switch frequently, which can significantly hurt user experience. Besides, the buffer occupancy fluctuates around a certain value (15s) for both clients. This is because in FESTIVE, the buffer occupancy threshold is preset to be 15s in the experiments. When the buffer occupancy of a client is higher than the threshold, the client will wait for some time before requesting the next segments. Otherwise, it will immediately send segment requests. Moreover, the threshold will be randomly adjusted when a segment is downloaded completely (before requesting the next segment), this also leads to small fluctuations of buffer occupancy. At 230s, the buffer occupancy decreases quickly and then increases to be around the threshold, this is mainly because the available bandwidth decreases dramatically from 3000 kbps to 1500 kbps at 230s.

We than analyze the performance of PANDA as shown in Fig. \ref{fig:rate_panda} and Fig. \ref{fig:buffer_panda}. Considering that the estimated bandwidth generally differs from (higher than most of the time) the fairly-shared bandwidth, the AIMD-based probing scheme is adopted to probe the fairly-shared bandwidth. As a result, a PANDA client switches its video bit-rate according to the probed bandwidth and its buffer occupancy. However, due to the characteristics of AIMD, such as slow start and frequent fluctuations, when the bandwidth changes, it takes long time to track the bandwidth well as Fig. \ref{fig:rate_panda} shows. For example, when client-2 joins the system at 50s, it takes more than 50s to probe the bandwidth well. Also, when client-2 joins the system, client-1 does not detect that the the fair-shared bandwidth is decreasing quickly. Until longer than 50s later, both clients have nearly the same probed bandwidth. Compared with FESTIVE, we can find that, with the probing scheme, PANDA achieves much smoother video bit-rate for both clients. However, since the probed bandwidth is used as the main signal for rate adaptation in PANDA, the video bit-rate also fluctuates to match the probed bandwidth. Besides, as the video bit-rate is discrete, the video bit-rate selected is usually lower than the probed bandwidth to ensure continuous video playback, which leads to higher buffer occupancy as shown in Fig. \ref{fig:buffer_panda}. However, this also leads to low bandwidth utilization efficiency since the video bit-rate is generally lower than the bandwidth.

Fig. \ref{fig:rate_our} and Fig. \ref{fig:buffer_our} show the performance of TFDASH. Similar to PANDA, the probing scheme is also adopted and the probed bandwidth is used to guide the rate adaptation. However, there are several major differences from PANDA. First, the proposed LIMD scheme tracks the bandwidth much better and quicker than AIMD as illustrated in Fig. \ref{fig:rate_our}. From Fig. \ref{fig:rate_our} we can find that when client-2 joins the system, it takes only about 10s to track the fair-share bandwidth well, and client-1 detects the bandwidth change quickly such that the probed bandwidths for both clients converge soon. Another major difference is that, instead of solely relying on the probed bandwidth for rate adaptation, the proposed dual-threshold buffer model effectively avoids buffer overflow/underflow. Besides, the proposed probability driven rate control logic can adequately break out the balance in case that the bandwidth is in perfect-subscription with unfair bandwidth sharing. Thanks to the dual-threshold buffer model, we smooth out bitrate by relaxing the smoothness of buffer occupancy provided that no buffer underflow (playback freeze) happens. This is reasonable since from the control system point of view, there is a fundamental conflict between maintaining stable video bitrate and maintaining stable buffer occupancy, due to the unavoidable network bandwidth variations. Nevertheless, from the end user point of view, video bitrate fluctuations are much more annoying than buffer occupancy oscillations. Besides, our rate adaptation scheme takes into account several factors that most influence the quality of experience, including buffer occupancy, video playback quality, video bit-rate switching frequency and amplitude. Therefore, the video bit-rate in Fig. \ref{fig:rate_our} is much smoother (more stable) than that with FESTIVE and PANDA. Besides, TFDASH allows the bit-rate to be higher than the probed bandwidth, thereby achieving significantly a higher bandwidth utilization efficiency and higher average video bit-rate compared with FESTIVE and PANDA under the some network conditions. At last, the buffer occupancy results in Fig. \ref{fig:buffer_our} shows that no buffer overflow/underflow happens, i.e., continuous video playback is guaranteed. From all the results, we can conclude that compared with FESTIVE and PANDA, TFDASH achieves much smoother and higher video bit-rate, and higher efficiency and stability.

Moreover, for the short-term bandwidth spikes, FESTIVE and PANDA generally switch up/down their bitrates, which is often unnecessary since there is enough buffered media/buffer space to accommodate the short-term bandwidth variations. In contrast, TFDASH can tolerate these short-term bandwidth variations more adequately, thanks to the proposed dual-threshold buffer model. This demonstrates the high robustness of TFDASH to short-term network variations.

\begin{figure*}[ht]
\centering
\subfigure[Inefficiency]{
\label{fig:inefficiency_2clients}
\includegraphics[width=0.3\textwidth]{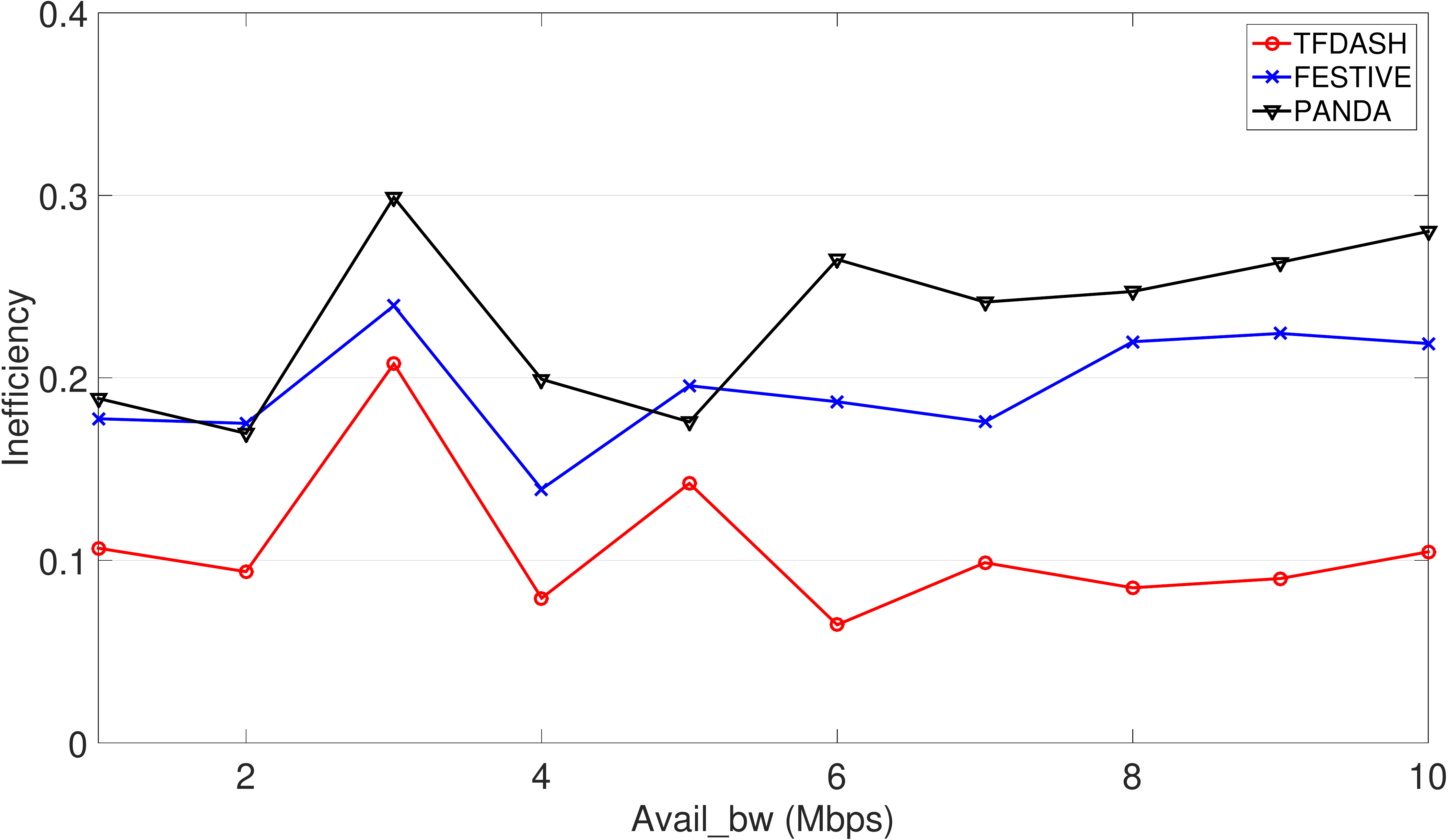}}
\hspace{0cm}
\subfigure[Instability]{
\label{fig:instability_2clients}
\includegraphics[width=0.3\textwidth]{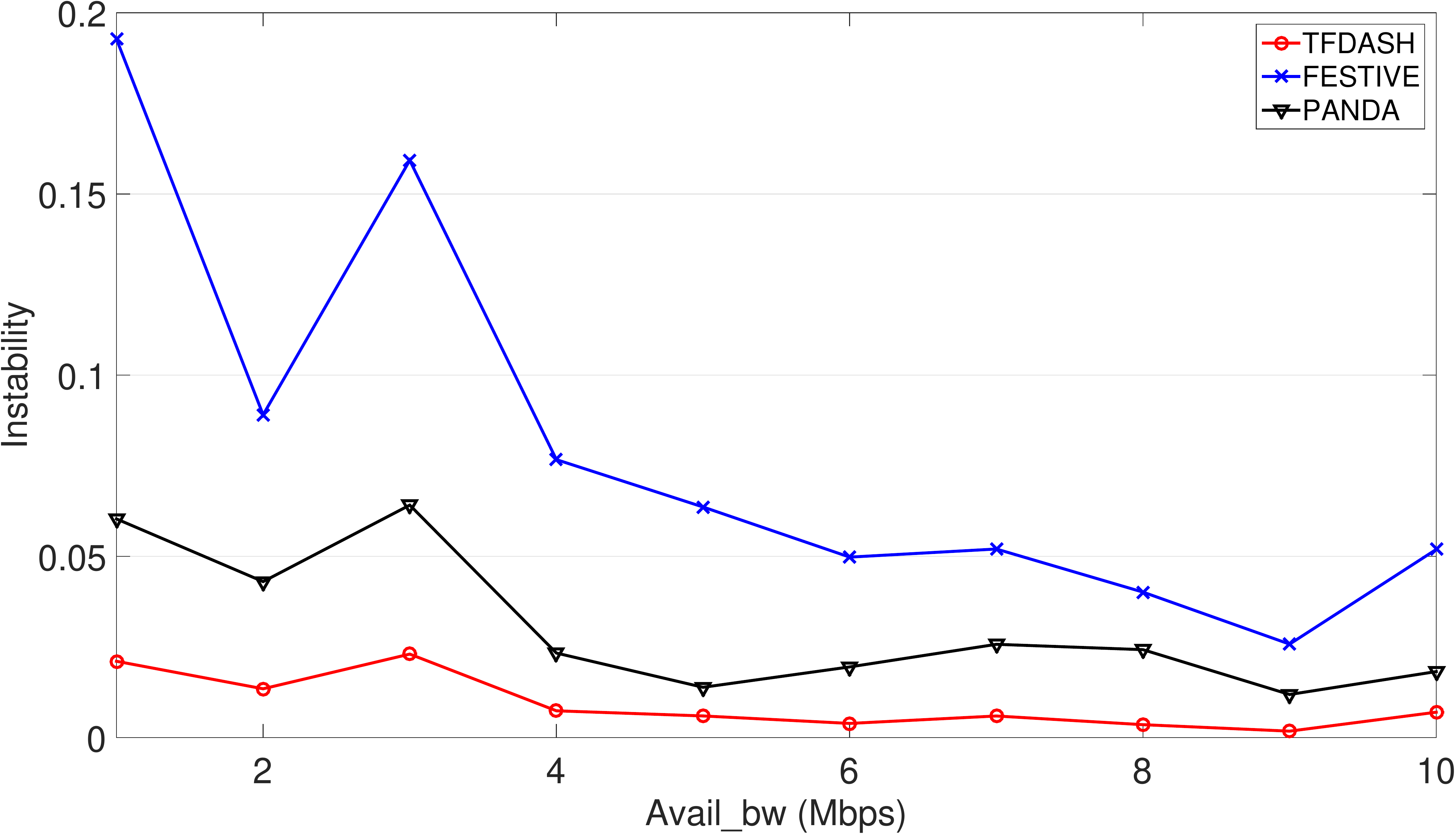}}
\hspace{0cm}
\subfigure[Unfairness]{
\label{fig:unfairness_2clients}
\includegraphics[width=0.3\textwidth]{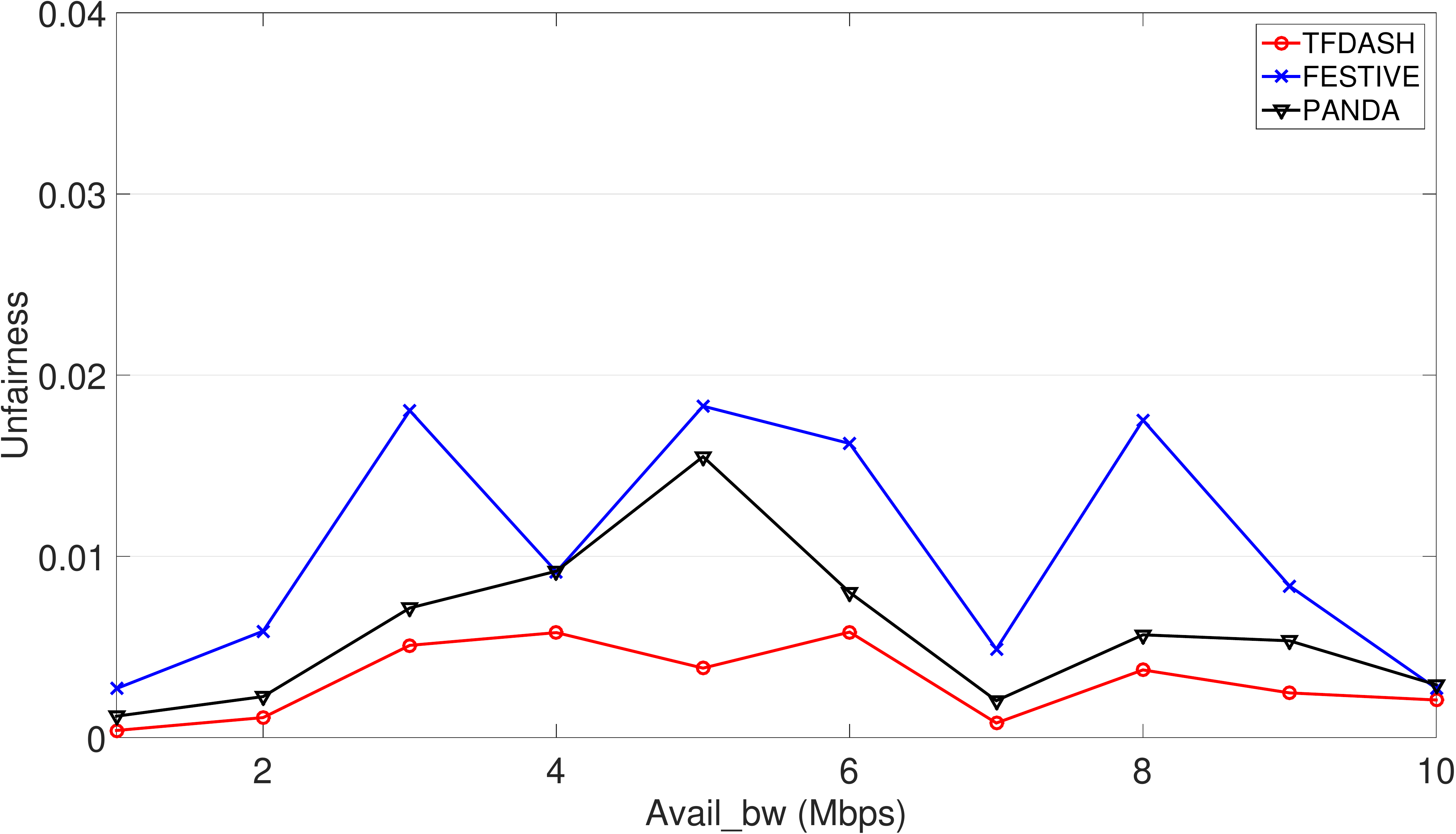}}
\caption{Performance evaluation with two competing clients when the available bandwidth changes from 1Mbps to 10Mbps.}
\hspace{0cm}
\label{fig:2clients}
\end{figure*}

\begin{figure*}[ht]
\centering
\subfigure[Inefficiency]{
\label{fig:inefficiency_10m}
\includegraphics[width=0.3\textwidth]{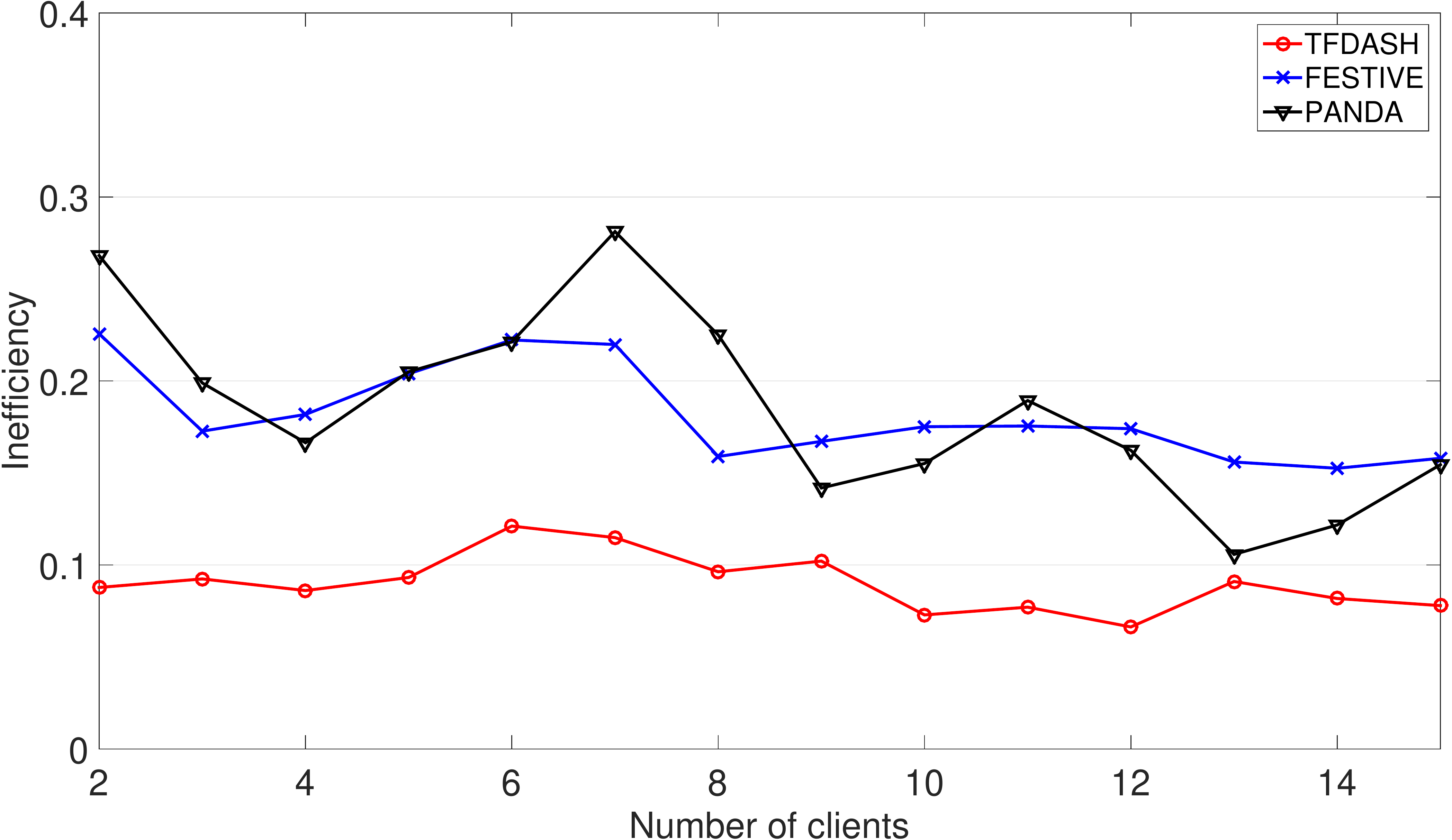}}
\hspace{0cm}
\subfigure[Instability]{
\label{fig:instability_10m}
\includegraphics[width=0.3\textwidth]{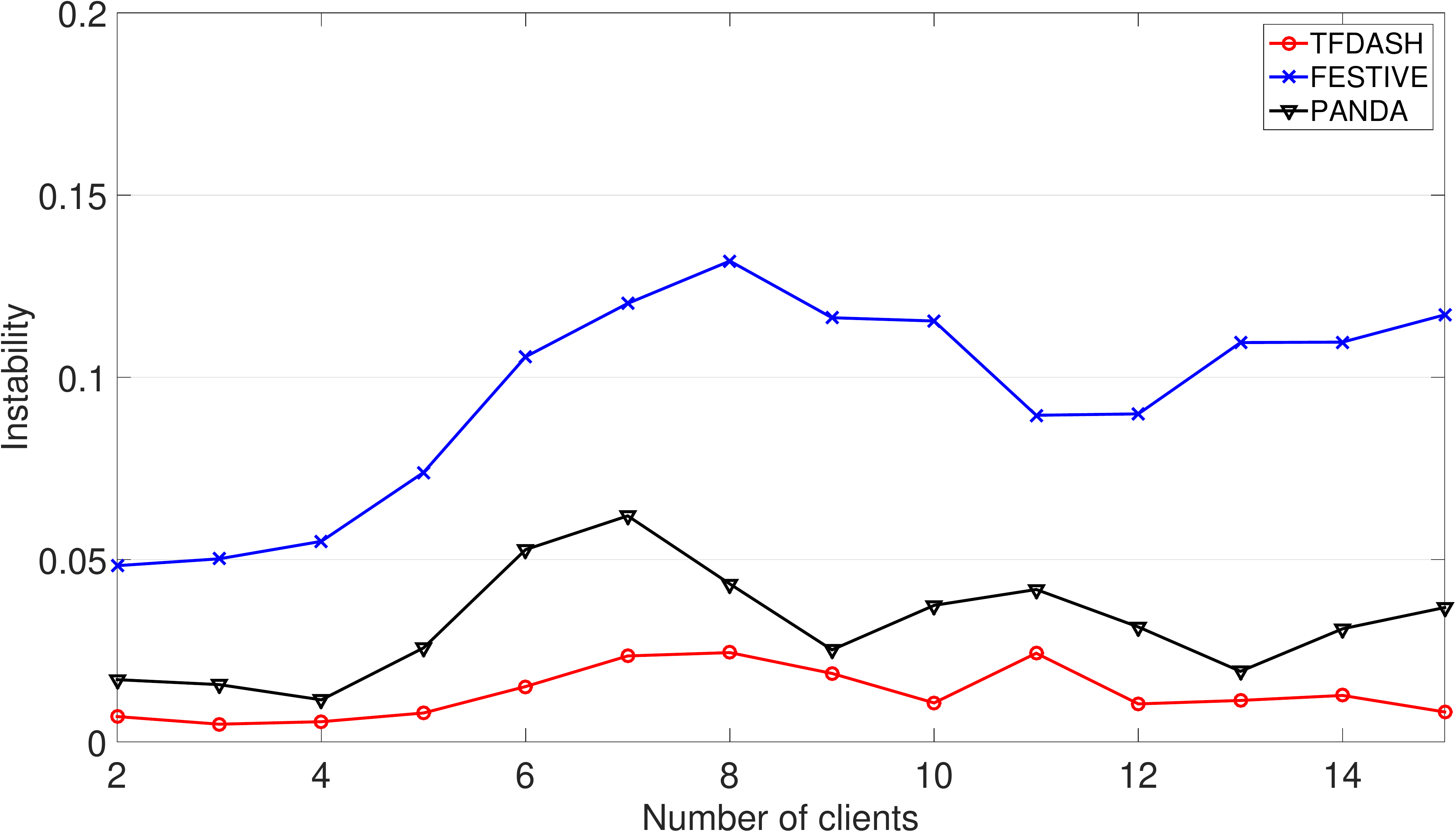}}
\hspace{0cm}
\subfigure[Unfairness]{
\label{fig:unfairness_10m}
\includegraphics[width=0.3\textwidth]{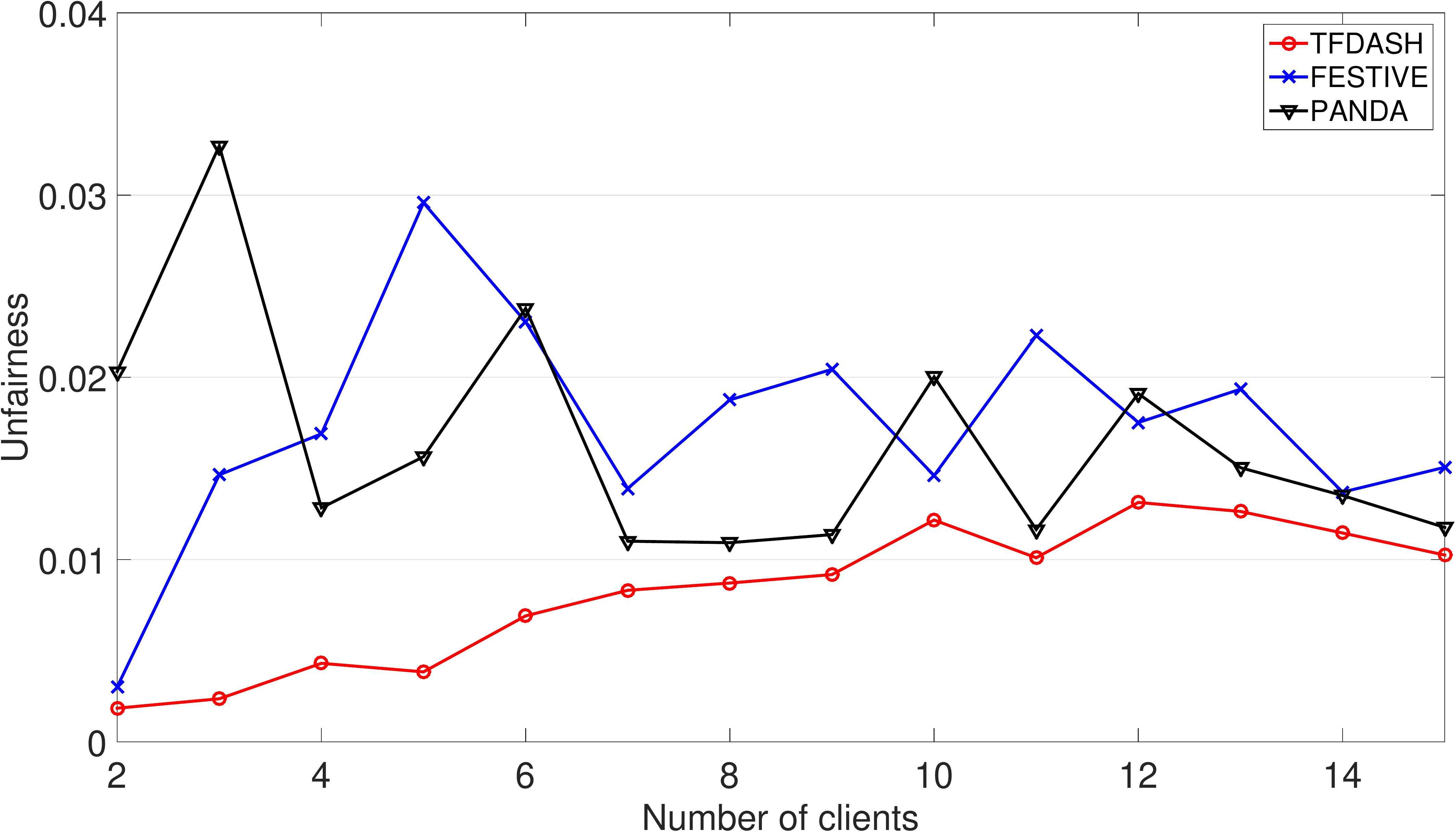}}
\caption{Performance evaluation when the available bandwidth is fixed at 10Mbps, and the number of competing clients changes from 2 to 15.}
\hspace{0cm}
\label{fig:10m}
\end{figure*}

\begin{figure*}[ht]
\centering
\subfigure[Inefficiency]{
\label{fig:inefficiency_1mper}
\includegraphics[width=0.3\textwidth]{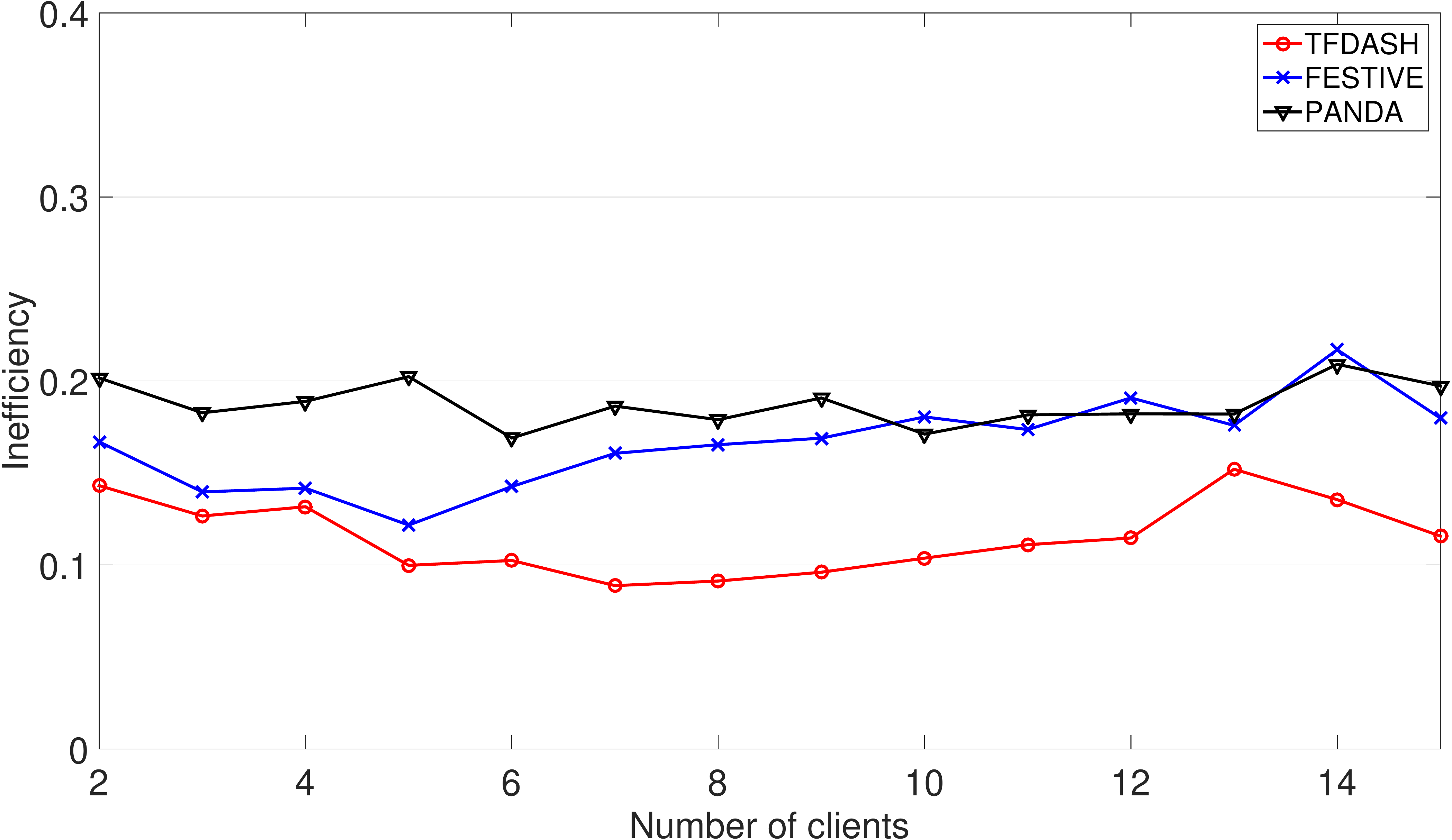}}
\hspace{0cm}
\subfigure[Instability]{
\label{fig:instability_1mper}
\includegraphics[width=0.3\textwidth]{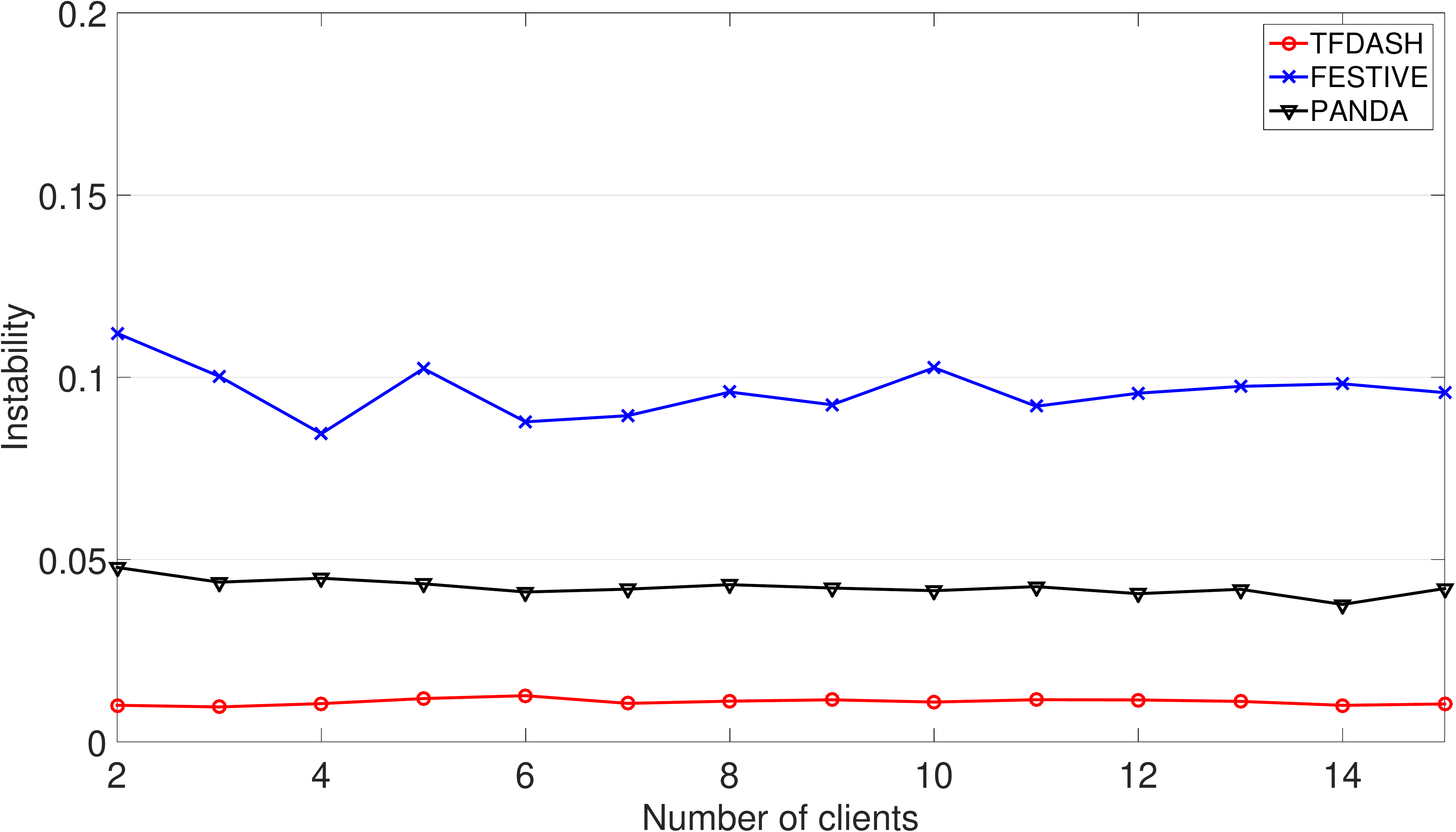}}
\hspace{0cm}
\subfigure[Unfairness]{
\label{fig:unfairness_1mper}
\includegraphics[width=0.3\textwidth]{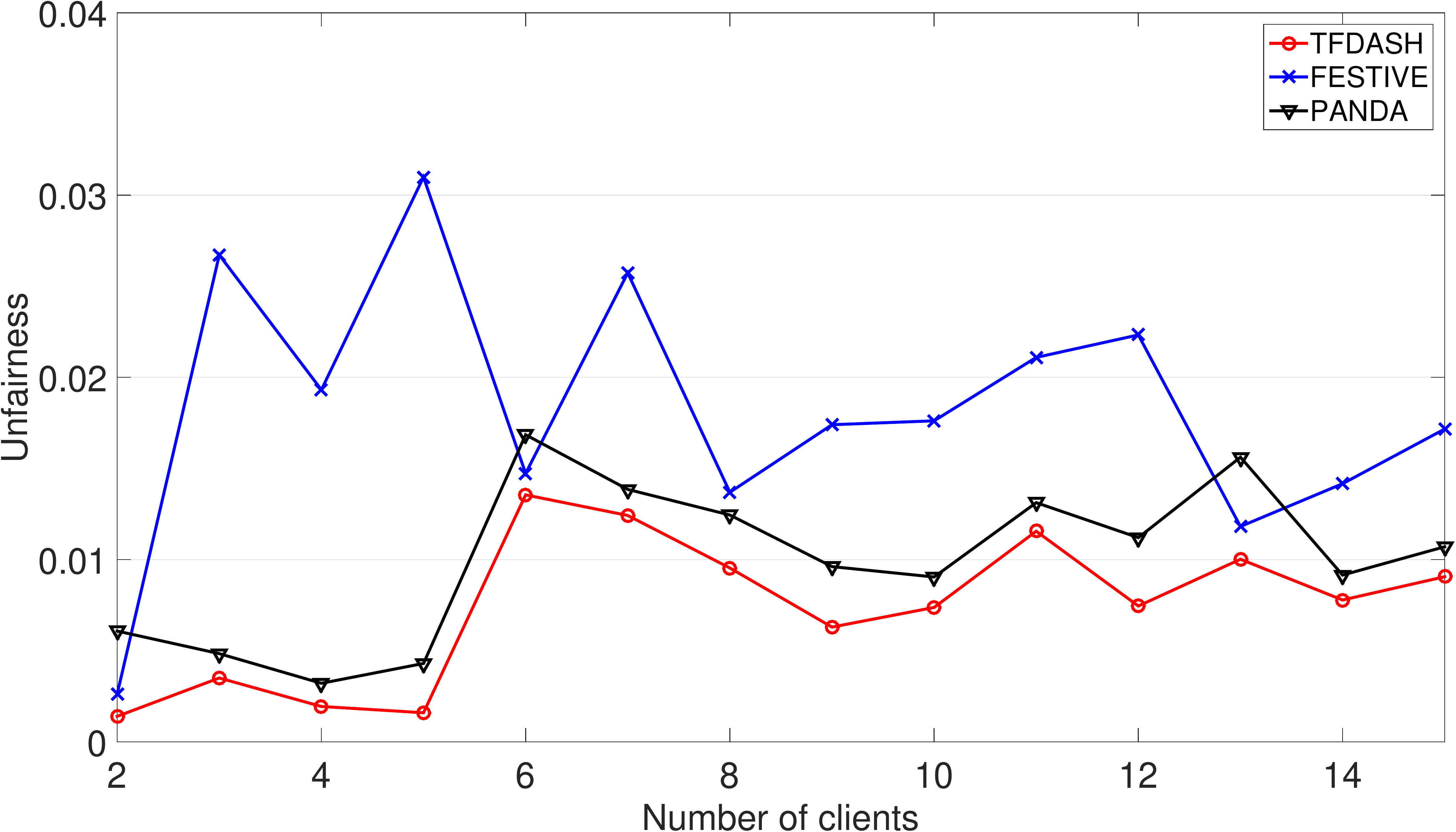}}
\caption{Performance evaluation under the condition that the average bandwidth for all the clients is 1Mbps.}
\hspace{0cm}
\label{fig:1mper}
\end{figure*}

\begin{figure*}[ht]
\centering
\subfigure[Inefficiency]{
\label{fig:inefficiency_rmcat}
\includegraphics[width=0.3\textwidth]{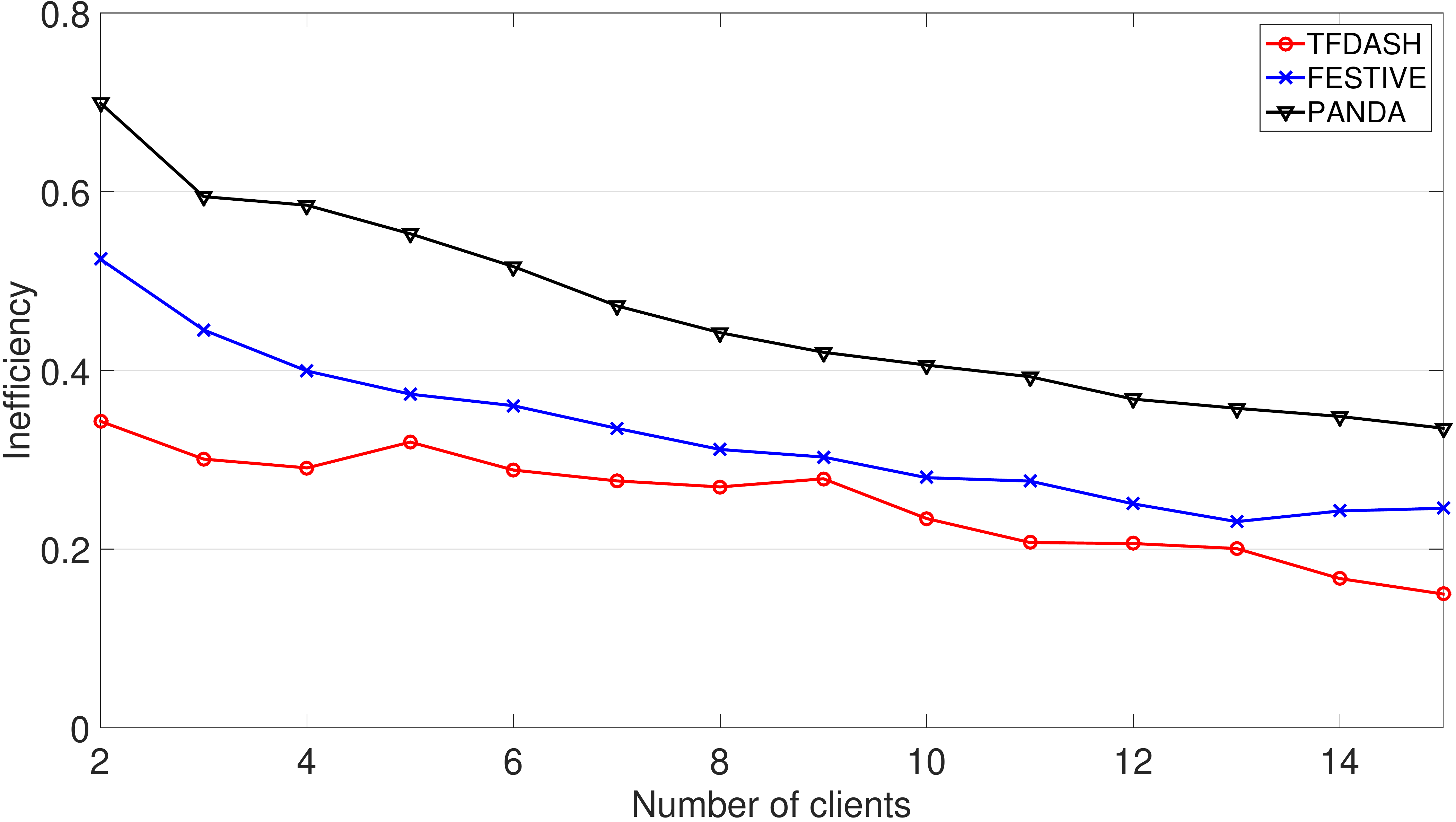}}
\hspace{0cm}
\subfigure[Instability]{
\label{fig:instability_rmcat}
\includegraphics[width=0.3\textwidth]{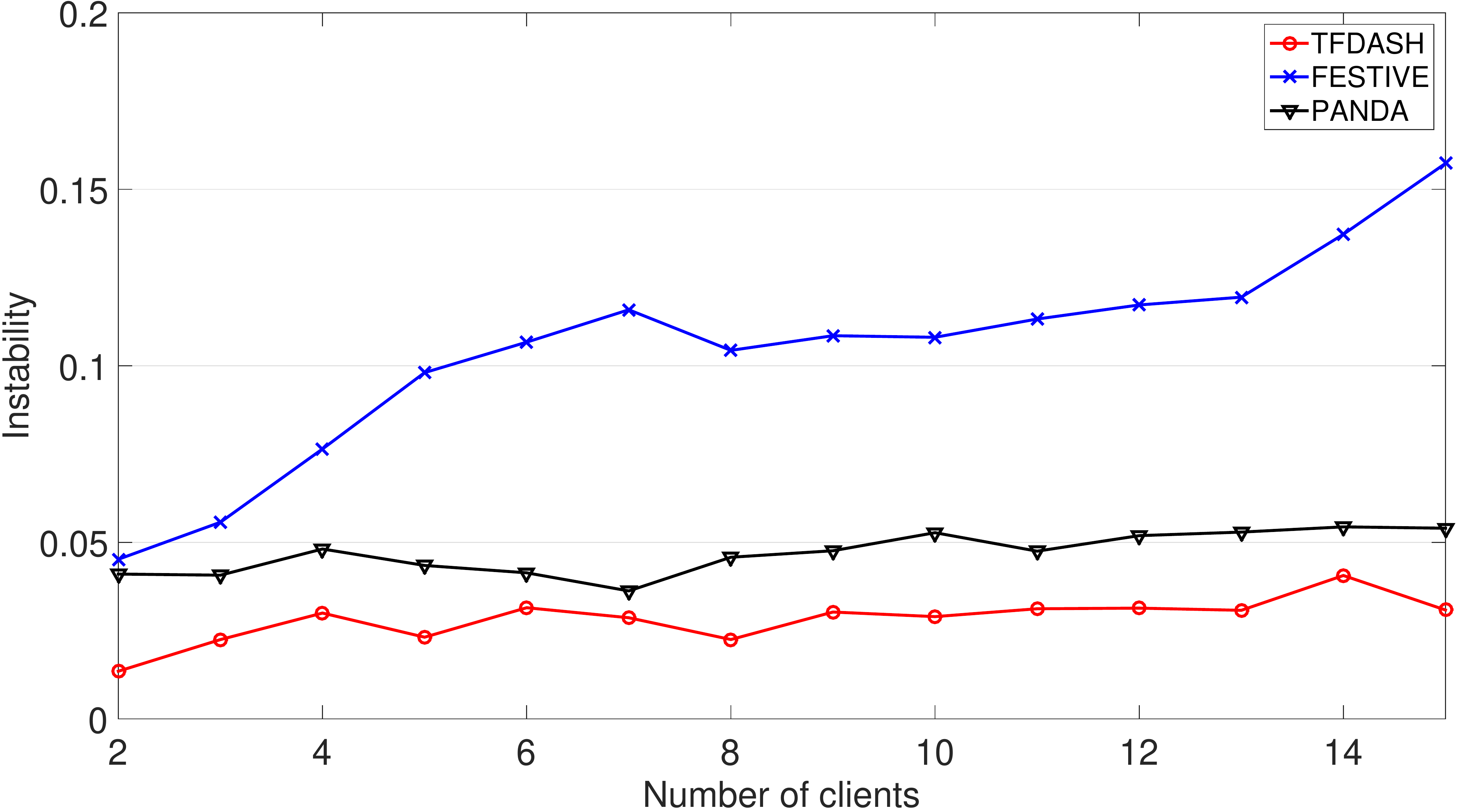}}
\hspace{0cm}
\subfigure[Unfairness]{
\label{fig:unfairness_rmcat}
\includegraphics[width=0.3\textwidth]{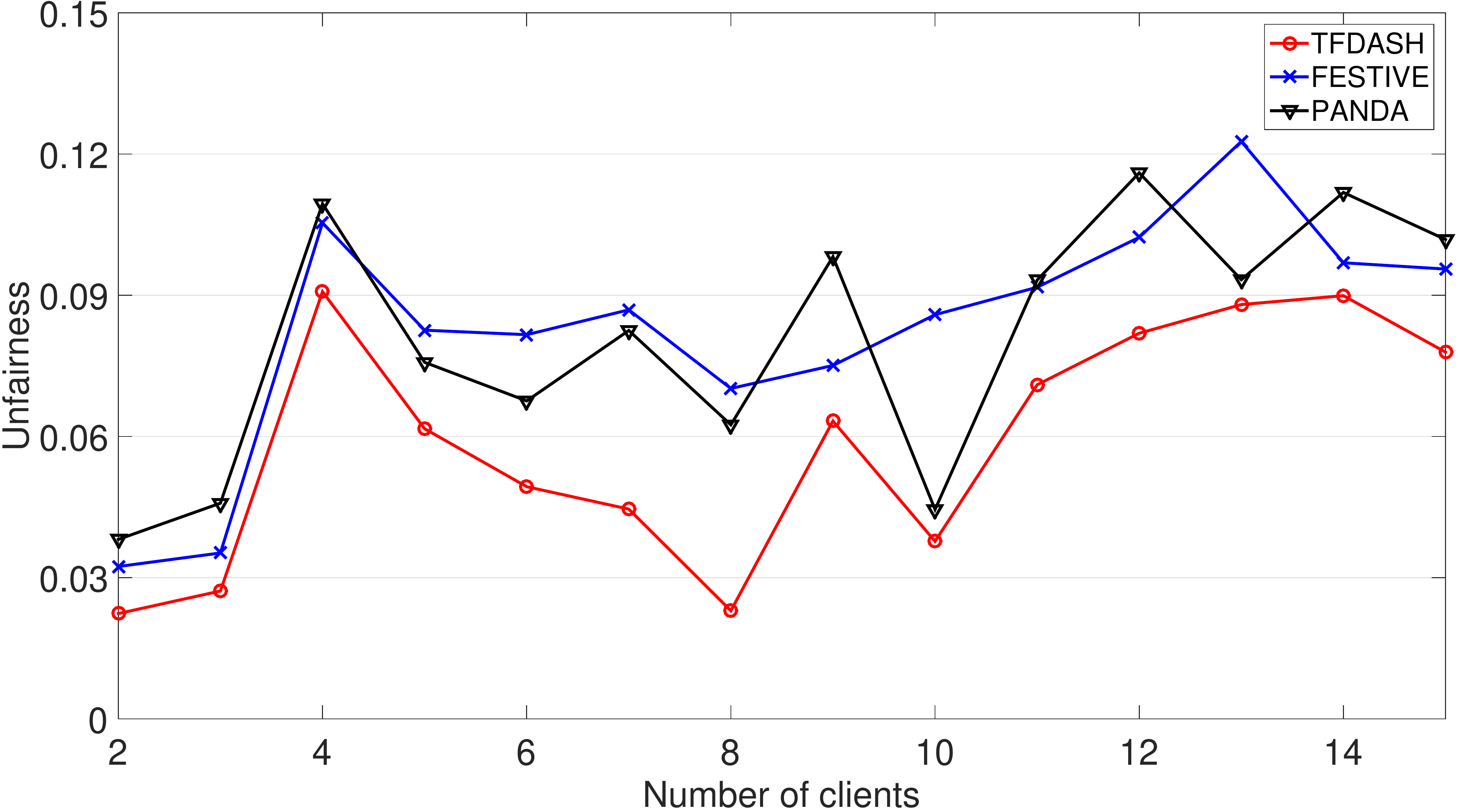}}
\caption{Performance evaluation with bottleneck capacity specified in Table \ref{tab:RMCAT}.}
\hspace{0cm}
\label{fig:rmcat}
\end{figure*}

\begin{table}[H]
\begin{center}
\caption{Path capacity variation pattern} \label{tab:RMCAT}
\begin{tabular}{|c|c|c|}
\hline
Variation pattern & Start time (s) & Path capacity (Mbps) \\
\hline
One & 0 & 8 \\
\hline
Two & 25 & 4 \\
\hline
Three & 50 & 7 \\
\hline
Four & 75 & 2 \\
\hline
Five & 100 & 4 \\
\hline
\end{tabular}
\end{center}
\end{table}

\subsection{Efficiency, Fairness ,and Stability Performances}
In this subsection, we compare the efficiency, stability, and fairness performances of all three schemes based on the following metrics \cite{jiang2014improving}:
\begin{itemize}
  \item \emph{Inefficiency}: The inefficiency at time is measured by $\left| {\frac{{\sum\limits_i {{v_{k,i}}} }}{b}} \right|$, where ${{v_{k,i}}}$ is the video bit-rate of the $k^{th}$ segment for client $i$, and $b$ is the available bandwidth. A value close to zero implies that the clients in aggregate are using as high an average bit rate as possible to improve user experience.
  \item \emph{Instability}: Recent studies suggest that users are likely to be sensitive to frequent and significant bit-rate switches \cite{mok2011inferring}. We define the instability metric as $\frac{{\sum\nolimits_{d = 0}^{{d_0} - 1} {\left| {{v_{k,k - d}} - {v_{k,k - d - 1}}} \right|} w\left( d \right)}}{{\sum\nolimits_{d = 1}^{{d_0}} {{v_{k,k - d}}} w\left( d \right)}}$, which is the weighted sum of all switch steps observed within the last $d_0 = 10$ segments divided by the weighted sum of bit-rates in the last $d_0$ segments. We use the weight function $w(d) = k-d$ to add linear penalty to more recent bit-rate switch.
  \item \emph{Unfairness}: The unfairness at $t$ is defined as $\sqrt {1 - JainFair_t} $, where $JainFair_t$ the Jain fairness index \cite{Jain} calculated on the rates $v_{k,i}$ at time $t$ over all clients.
\end{itemize}

We first compare the performance of all schemes with two competing clients, and the avail\_bw varies from 1 Mbps to 10 Mbps. Fig. \ref{fig:2clients} show that TFDASH always achieves the lowest inefficiency, instability, and unfairness, i.e., compared with FESTIVE and PANDA, TFDASH can provide higher and smoother video bit-rate, and besides, it can also better guarantee the fairness between the clients. As the avail\_bw increases, the performance does not monotonically increase or decrease. This is mainly because the video bit-rate is discrete so that for a given avail\_bw, if the fair-share bandwidth is equal (close) to an available video bit-rate, generally better performance can be achieved, and vice versa.

Moreover, we also vary the number of competing clients from two to fifteen to evaluate the performance of the three schemes with avail\_bw constraint of 10 Mbps as shown in Fig. \ref{fig:10m}. Similar to the results in Fig. \ref{fig:2clients}, TFDASH always achieve the best performance in terms of efficiency, stability, and fairness. The performance improvement mainly come from the LIMD based bandwidth probing scheme, the proposed dual-threshold buffer model, and the probability-driven rate control logic. 

Then, we vary both the number of competing clients and avail\_bw to evaluate the performances of all schemes. Specifically, the fairly shared bandwidth for each client is kept to be 1 Mbps, i.e., the avail\_bw is set to be 2 Mbps for two competing clients, the avail\_bw is set to be 3 Mbps for three competing clients, and so on. The results in Fig. \ref{fig:1mper} show that TFDASH still achieves the best performance in all cases, demonstrating the high efficiency of TFDASH. As for the performance of efficiency and stability, Fig. \ref{fig:inefficiency_1mper} and Fig. \ref{fig:instability_1mper} show that in different cases, all the schemes have insignificant performance fluctuations, this is because the fair-share bandwidth plays an important role in the decision of video bit-rate, which is always the same (equal to 1 Mbps) in this experiment. While in terms of fairness, TFDASH performs the best, but all schemes fluctuate as the number of clients increase. This is obvious since fairness is easy to be affected by the number of competing clients.

At last, we conduct performance evaluation according to the suggested test cases for multiple clients in IETF RMCAT \cite{rmcat}, we conduct new experiments based on the path capacity variation pattern listed in Table \ref{tab:RMCAT} with a corresponding end time of 125 seconds. We use background non-adaptive UDP traffics to simulate a time-varying bottleneck for congestion controlled media flows. In the experiments, the physical path capacity is 10 Mbps and the UDP traffic source rate changes over time as $10-x$ Mbps, where $x$ is the bottleneck capacity specified in Table \ref{tab:RMCAT}.

As shown in Fig. \ref{fig:rmcat}, TFDASH always achieves the best performance in terms of efficiency, stability, and fairness. The performance improvement mainly comes from the LIMD based bandwidth probing scheme, the proposed dual-threshold buffer model based and probability driven rate control logic. Specifically, the LIMD-based bandwidth probing scheme can track the bandwidth significantly better and quicker, the proposed dual-threshold buffer model can achieve a good trade-off between the smoothness and high efficiency, and the probability-driven rate control logic can improve the fairness among clients competing for channel bandwidth.

\begin{figure}[ht]
\centering
\includegraphics[width=0.48\textwidth]{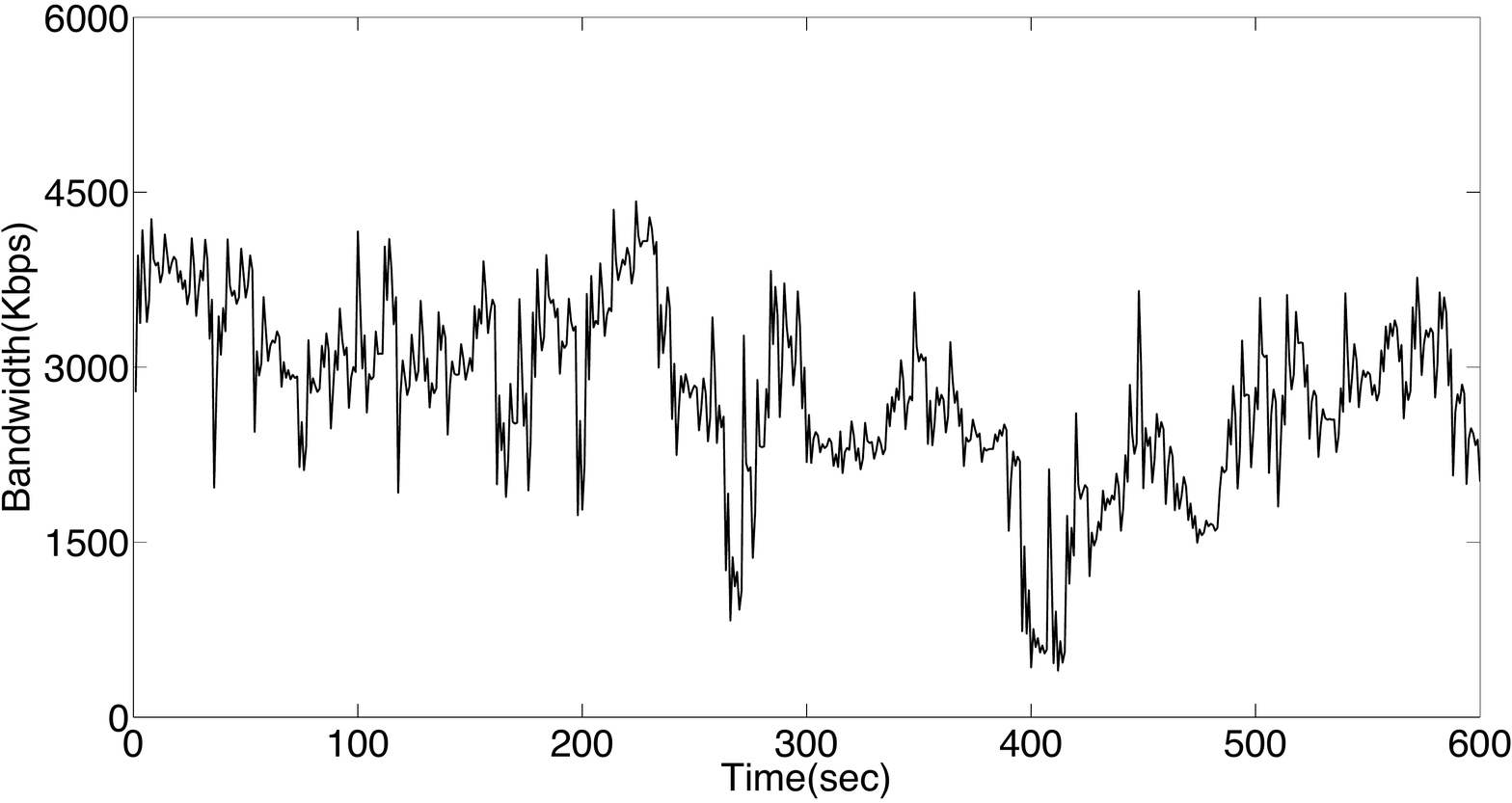}
\caption{Bandwidth trace collected from the PlanetLab with a server deployed in Hong Kong and clients deployed in Beijing, China.}
\label{fig:trace}
\end{figure}

\begin{table}[H]
\caption{Performance Summary Under Internet Bandwidth Trace}
\label{tab:trace}
\begin{center}
\begin{tabular}{ c | p{0.065\textwidth}<{\centering}| p{0.065\textwidth}<{\centering}| p{0.065\textwidth}<{\centering}| p{0.065\textwidth}<{\centering}}\hline \hline
Metrics & Two FESTIVE clients & Two PANDA clients & Two TFDASH clients &All six DASH clients \\ \hline
Inefficiency &  0.278 & 0.289 & 0.134 & 0.305  \\  \hline
Instability & 0.187 & 0.093 & 0.031 & 0.256    \\  \hline
Unfairness & 0.035 & 0.024 & 0.012 & 0.067    \\  \hline
\end{tabular}
\end{center}
\end{table}

\subsection{Performance with three types of competing clients}
In DASH, it is important to design an appropriate rate adaptation algorithm so that multiple clients can fairly share bandwidth with high stability and efficiency. However, in fact all the clients adapt their bitrate independently in the following aspects:  i) for each client, the bitrate is adapted based on its own rate adaptation logic, ii) no information is exchanged among the clients, iii) the algorithm is distributedly executed in each client and no centralized controller/server is needed. However, since no DASH algorithms has been widely deployed, different in-house algorithms are used in commercial products from different companies, such as Apple?s HLS, Netflix, and YouTube. Thus, it is hard for a newly designed algorithm to achieve high performance, especially in terms of fairness performance, when competing with different DASH clients implementing different rate adaptation algorithms. 

In this section, we consider the scenario with different types of clients competing bandwidth over a bottle-neck link. In the experiment, we use a bandwidth trace collected from real networks  which contains more variations as shown in Fig. \ref{fig:trace} shows. The trace is collected from the Plantlab with a server physically deployed in Hong Kong and clients deployed in Beijing, China. 

Considering that the bandwidth is not high enough to support too many clients, we implemented two TFDASH clients, two FESTIVE clients, and two PANDA clients at the same time. The six DASH clients compete the bandwidth based on their own rate adaptation algorithms and the results are show in Table \ref{tab:trace}. The results show that TFDASH achieves better performance than FESTIVE and PANDA in terms of fairness, stability, and efficiency. However, compared with the results in Sec. VI.C, all types of clients perform a bit worse than the scenario with only the same type of clients. Moreover, when comparing the overall performance of all six clients, we can observe that all metrics (fairness, stability, and efficiency) become worse, this is mainly because different types of clients are designed with different objectives without considering the other types of clients, making one type of clients difficult to cooperate with others fairly and efficiently.

\section{Conclusion}
In this paper, we addressed the rate adaptation problem with multiple DASH clients competing bandwidth over a bottle-neck link by considering the efficiency, stability, and fairness among clients. We proposed a throughput-friendly DASH client to intelligently and dynamically switch the video bit-rate so as to reach a good trade-off among these objectives. Specifically, we proposed a Logarithmic Increase Multiplicative Decrease (LIMD) based bandwidth probing scheme to guide the rate adaptation, by which the fair shared bandwidth can be effectively and quickly detected. Besides, we also proposed a dual-threshold based rate adaptation scheme to guarantee continuous video playback with high visual quality. Moreover, the proposed probability driven rate adaption logic further considers several key factors including buffer occupancy, video playback quality, video bit-rate switching frequency and amplitude to achieve the high visual video quality with good fairness among the competing clients. Our experiments demonstrate the high efficiency and good performance of our proposed TFDASH compared with state-of-the-art schemes.

\begin{IEEEbiography}[{\includegraphics[width=1in,height=1.25in,clip,keepaspectratio]{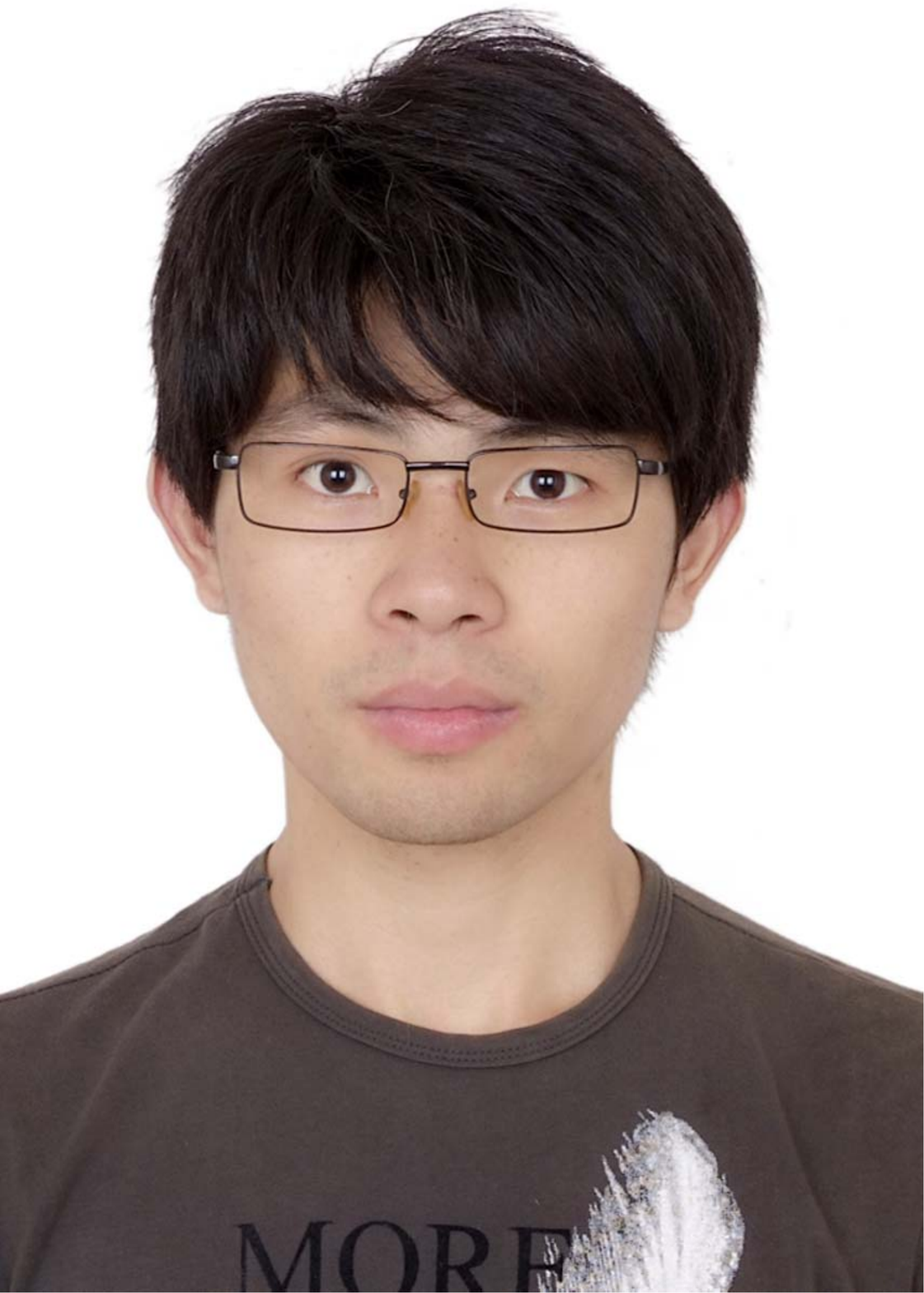}}]{Chao Zhou} received his Ph.D. degree from the Institute of Computer Science \& Technology, Peking University, Beijing, China, in 2014.

  He has been with Beijing Kuaishou Technology Co., Ltd. as a Senior Algorithm engineer. Before joining Kuaishou, he was a Senior Research Engineer with the Media Technology Lab, CRI, Huawei Technologies CO., LTD, Beijing, China.
  
  Dr. Zhou's research interests include HTTP video streaming, joint source-channel coding, and multimedia communications and processing. He has been the reviewer for \textsc{IEEE Transactions on Circuits and Systems for Video Technology}, \textsc{IEEE Transactions on Multimedia}, \textsc{IEEE Transactions on Wireless Communication} and so on. He received Best Paper Award presented by IEEE VCIP 2015, and Best Student Paper Awards presented by IEEE VCIP 2012.

\end{IEEEbiography}

\begin{IEEEbiography}[{\includegraphics[width=1in,height=1.25in,clip,keepaspectratio]{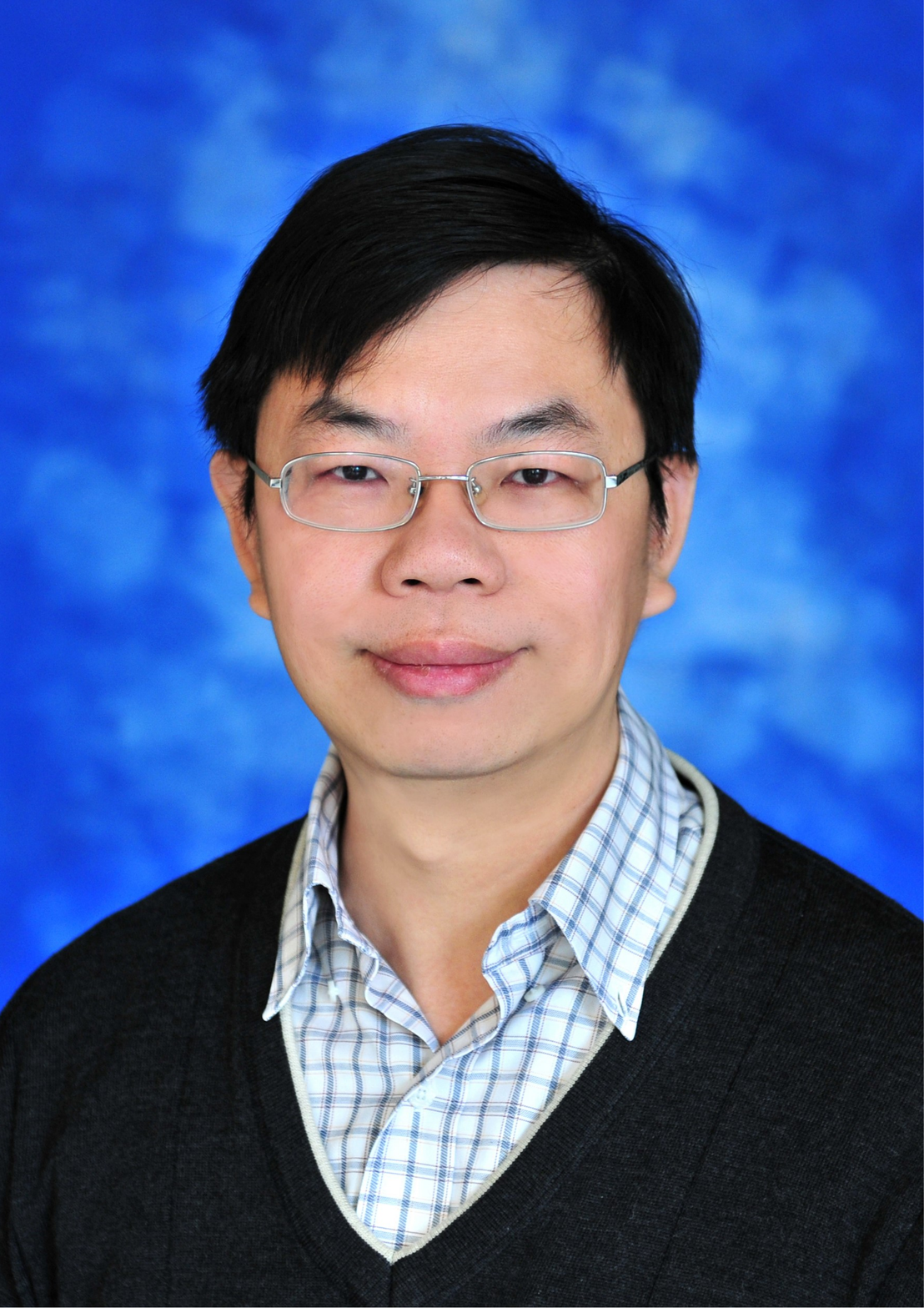}}]{Chia-Wen Lin}
	(S'94-M'00-SM'04) received his Ph.D. degree in electrical engineering from National Tsing Hua University (NTHU), Hsinchu, Taiwan, in 2000.
	
	He is currently Professor with the Department of Electrical Engineering and the Institute of Communications Engineering, NTHU. He was with the Department of Computer Science and Information Engineering, National Chung Cheng University, Taiwan, during 2000--2007. Prior to joining academia, he worked for the Information and Communications Research Laboratories, Industrial Technology Research Institute, Hsinchu, Taiwan, during 1992--2000. His research interests include image and video processing and video networking.
	
	Dr. Lin has served as an Associate Editor of \textsc{IEEE Transactions on Image Processing}, \textsc{IEEE Transactions on Circuits and Systems for Video Technology}, \textsc{IEEE Transactions on Multimedia}, \textsc{IEEE Multimedia}, and Journal of Visual Communication and Image Representation. He was a Steering Committee member of \textsc{IEEE Transactions on Multimedia} from 2014 to 2015. He was Chair of the Multimedia Systems and Applications Technical Committee of the IEEE Circuits and Systems Society from 2013 to 2015. He served as Technical Program Co-Chair of IEEE ICME 2010, and will be the General Co-Chair of IEEE VCIP 2018 and Technical Program Co-Chair of IEEE ICIP 2019. His papers won Best Paper Award of IEEE VCIP 2015, Top 10\% Paper Awards of IEEE MMSP 2013, and Young Investigator Award of VCIP 2005. He received the Young Investigator Award presented by Ministry of Science and Technology, Taiwan, in 2006.	
\end{IEEEbiography}

\begin{IEEEbiography}[{\includegraphics[width=1in,height=1.25in,clip,keepaspectratio]{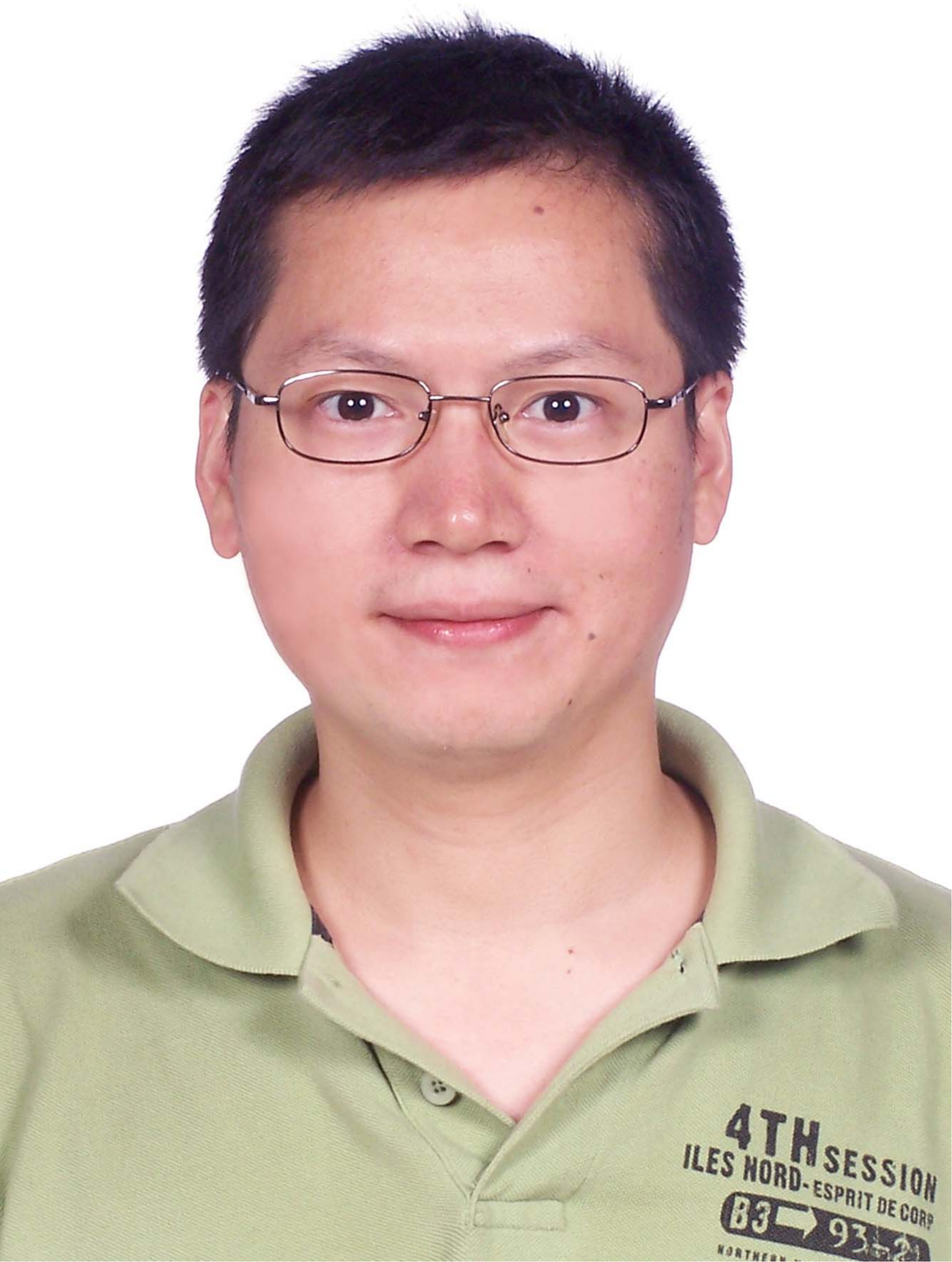}}]{Xinggong Zhang}
	(M'11) received the B.S. degree from Harbin Institute of Technology and the M.S. degree from Zhejiang University in 1995 and 1998 respectively, and the Ph.D degree from Peking University, Beijing, China, in 2011.
	He is currently an Associate Professor with the Institute of Computer Science and Technology, Peking University. He has been a senior Research Staff with Founder Research China. His general research interests lie in multimedia networking and video communications. His current research directions include video conferencing, dynamic HTTP streaming, and content-centric networking.
	Dr. Zhang was a recipient of the First Prize of the Ministry of Education Science \& Technology Progress Award in 2006, and the Second Prize of the National Science \& Technology Award in 2007.
\end{IEEEbiography}

\begin{IEEEbiography}[{\includegraphics[width=1in,height=1.25in,clip,keepaspectratio]{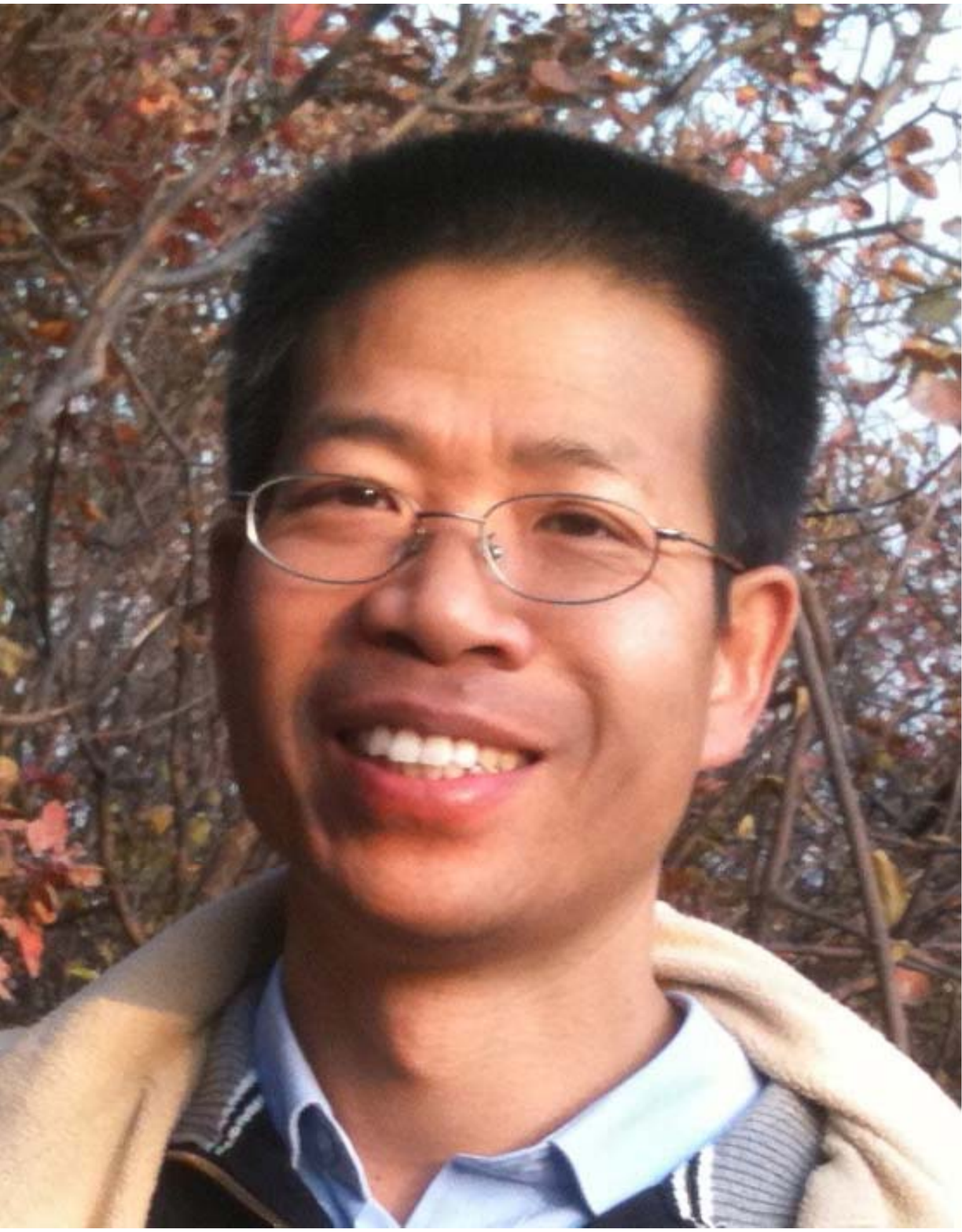}}]{Zongming Guo}
	received the B.Sc. degree in mathematics, the M.Sc. and the Ph. D degrees in computer science from Peking University, China, in 1987, 1990 and 1994, respectively.
	
	He is currently Dean of Institute of Computer Science \& Technology, Peking University. His current research interests include streaming media technology, IPTV and mobile multimedia, and image and video processing. He has published over 80 technical articles in refereed journals and conference proceedings in the areas of multimedia, image and video compression, image and video retrieval, and watermarking.
	
	Dr. Guo led the research \& developing team, which won the first prize of The State Administration of Radio Film and Television, the first prize of Ministry of Education Science and Technology Progress Award and the second prize of National Science and Technology Award in 2004, 2006 and 2007, respectively. He received Government Allowance granted by the State Council in 2009.
\end{IEEEbiography}

\end{document}